\documentclass[a4paper,amsmath,amssymb,pre]{revtex4-1}
\usepackage{graphicx}
\usepackage{rotating}

\usepackage{dcolumn}
\usepackage{color}
\usepackage{ulem}

\usepackage{bm}% bold math
\usepackage{latexsym}% bold math
\usepackage{epic,eepic}
\usepackage{subfigure}
\usepackage{overpic}

\def\r{\mathbf{r}}
\def\k{\mathbf{k}}

\def\ka{Kob-Andersen}
\def\bm{Biroli-M\'ezard}

\begin{document}

\author{Davide Cellai}
% \email{davide.cellai@ucd.ie}
% \affiliation{School of Computer Science and Informatics, University College Dublin, Belfield, Dublin 4, Ireland}
\altaffiliation[]{This work was carried out while DC was at the School of Chemistry and Chemical Biology, University College Dublin,
Belfield, Dublin 4, Ireland.}
\affiliation{Department of Mathematics and Statistics, University of Limerick, Limerick, Ireland}

\author{Andrzej Z.\ Fima, Aonghus Lawlor, and Kenneth A.\ Dawson}
\affiliation{Centre For BioNano Interactions (CBNI),
School of Chemistry and Chemical Biology, University College Dublin,
Belfield, Dublin 4, Ireland}

\title{Lattice Model of Glasses}

\begin{abstract}
Glass-forming liquids have been extensively studied in recent decades, but there is still no theory that fully describes these systems, and the diversity of treatments is in itself a barrier to understanding. Here we introduce a new simple model that (possessing both liquid-crystal and glass transition) unifies different approaches, producing  most of the phenomena associated with real glasses, without loss of the simplicity that theorists require. Within the model we calculate energy relaxation, non-exponential slowing phenomena, the Kauzmann temperature and other classical signatures.
Moreover, the model reproduces a subdiffusive exponent observed in experiments of dense systems.
The simplicity of the model allows us to identify the  microscopic origin of glassification, leaving open the possibility for theorists to make further progress.
\end{abstract}

\date{\today}
\maketitle

\section{Introduction}

Understanding the nature of the glass transition is considered to be one of the great outstanding problems of the
condensed state of matter \cite{angell2000,binder2005}.
Advances have been real, but progressive, and there is still no theory of the
glassy state that explains all of its properties.  Quite different approaches and levels of description have arisen within the
field, each with acknowledged strengths. However, the diversity of levels of description has in itself become a barrier to
progress and understanding, for in many cases there is no apparent connection between the different strands of research.

Here we cannot summarize the numerous efforts, but mention only several trends in the field that are relevant to this paper.
The glass problem is related to a wider set of phenomena, collectively named \emph{dynamical arrest}: that process in which many particles dramatically slow in a concerted manner
\cite{kob1993,pal2008,jackle1994,mezard2000,toninelli2006,liu1998,sellitto2006,mehta2000,biroli2002,bouchaud2003,garrahan2002,ritort2003,starr2002}.
One interpretation is as follows: for simple repulsive interactions, with increasing density, progressive loss of space around
a typical particle leads it to become effectively trapped by its neighbors, a phenomenon often termed \emph{caging}
\cite{gotze1991,giovambattista2003}.
Occasionally, the particle can escape from the cage and make longer movements before being trapped in another cage.
Kinetically constrained models, such as the one introduced by Kob and Andersen \cite{kob1993}, represent the intra-cage behavior of glassy systems on a lattice and produce blocked nonergodic states and dynamical heterogeneities \cite{lawlor2005,degregorio2005}.

Despite their success, such simple models are criticized because they  have no energy relaxation, possess no underlying crystal phase, and fail to exhibit the correct decay of dynamical correlations with time.
Experimental studies of glass transitions \cite{angell2000} are primarily presented with temperature as the
control parameter, and the heat capacity plays a central role \cite{adam1965,binder2005}.
While continuum calculations reflect many of these aspects of the system rather well \cite{sastry1998,giovambattista2003,moreno2006}, computer simulations of such accurate models are enormously challenging to carry out and to interpret in terms of simple
theories.

The aim of this paper is to bridge the gap between these two poles of scientific study by presenting a treatment that has the simplicity of the lattice models but schematically behaves as a true glass.
The outcome of our work is a quite realistic model of the glass which produces a synthesis of the elements of the glass phenomenon.

In Section II the model is introduced and the meaning of its definition is illustrated, in Section III the main equilibrium properties of the model are presented, in Section IV and V the dynamics and the aging behavior are analyzed. Finally, in Section VI the conclusions of this work are presented.

\section{The model}
One of the aims of the model is to describe crystallization with  as simple a theoretical tool as possible.
Continuum models have had  extraordinary success in this area.
Remarkably, simulations of hard spheres show the full range of phenomena which have been observed in experiments \cite{hoover1968,rintoul1996b,sastry1998}.
Here we try to represent, in a lattice model, the behavior of a system of quasi-hard spheres in the continuum.
Our model consists of only two ingredients: a repulsive interaction described by a simple Hamiltonian and a kinetic rule affecting the probability of a movement.

\subsection{The Repulsive Hamiltonian}
Several Hamiltonian lattice models have been introduced to mimic on a lattice the relevant properties of hard spheres with short ranged interactions \cite{biroli2002,ciamarra2003b,cellai2004,marinari2006,witman2006,krzakala2008}.
The equilibrium properties of our model are inspired to the {\bm} model \cite{biroli2002}.
The {\bm} model is defined by a many body short ranged repulsion between nearest neighbors.
In their paper, Biroli and M\'ezard are mainly interested in studying the glassy behavior.
Therefore, they consider a mixture of two types of particle in order to hinder crystallization.
In out model, instead, as the dynamical slowing is governed by the kinetic rule, we use the {\bm} approach only to guarantee the presence of an underlying liquid-crystal transition at the equilibrium.
Therefore, for our purposes we consider the single-type particle case.
Moreover, in order to represent softness, we use a modified definition of the
Hamiltonian, which reads:
\begin{equation}
\label{eq:hamiltonian}
H = V_R \sum_{j=1}^{V}  \,(n_j - c_R)\, \theta(n_j - c_R)
\end{equation}
Here $c_R$ is the maximum number of nearest neighbors that may
surround a particle without it incurring an energy cost, $n_j$ the number of nearest neighbors of the particle at the $j$-th site, $V_R$ is the
strength of the repulsive interaction, $\theta(x)$ is the Heaviside function and $V$ the total number of sites (volume).
Of course, after this extension the model is no longer a-thermal.
In fact, at $T=0$ the hard repulsion is recovered as a particular case.

\begin{figure}[hbt]
	\begin{center}
		\includegraphics[width=0.95\columnwidth]{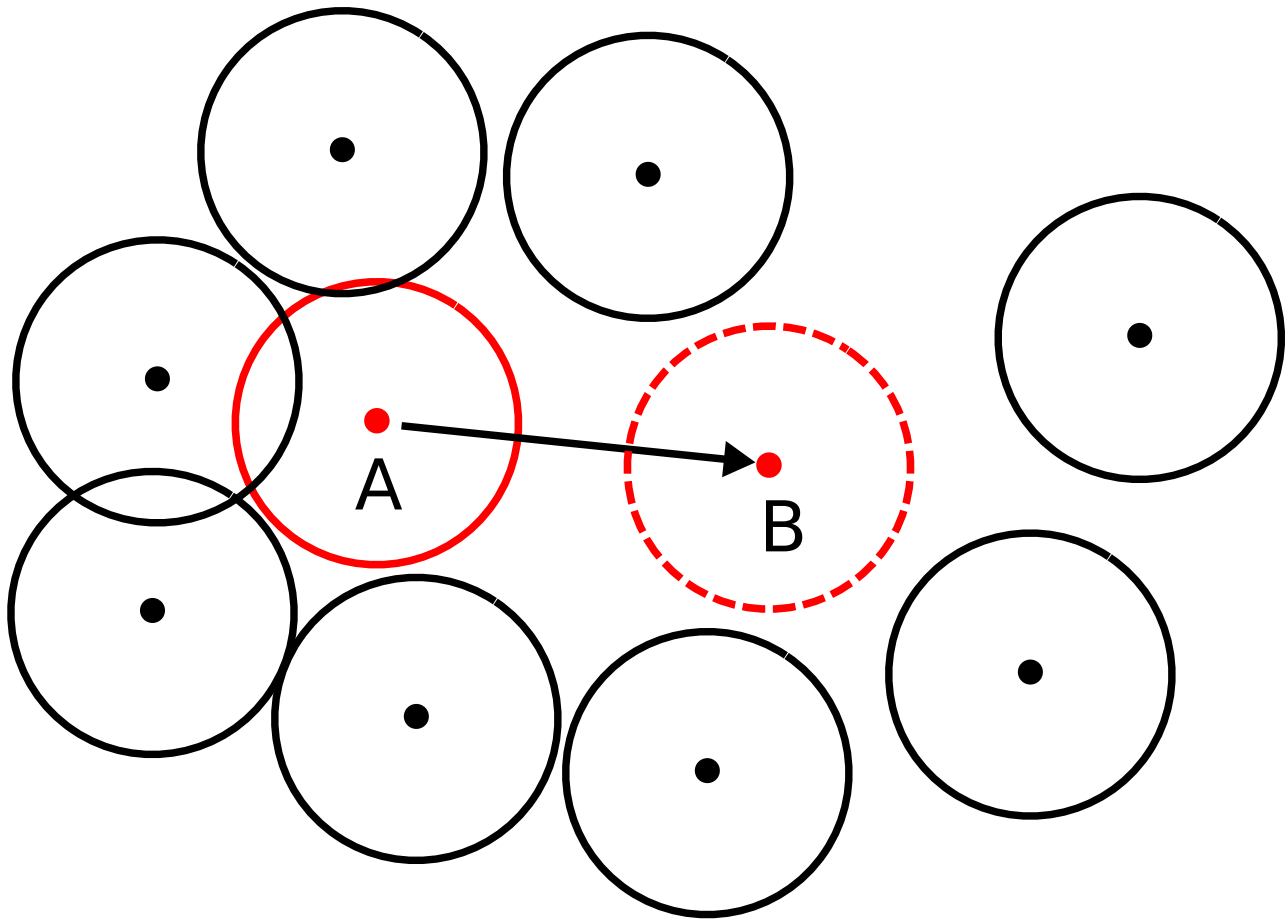}%
	\end{center}
\caption{Schematic drawing of soft disks in two dimensions.
The large contour of each particle represents an effective size, so that  soft particles can overlap.
For example, the particle in  position $A$ has higher energy than it would have in position $B$.
Therefore, the movement from position $A$ to $B$ is energetically favourable.}
\label{fig:meaning_softbm}
\end{figure}

The meaning of this Hamiltonian is conceptually illustrated in
Figure \ref{fig:meaning_softbm}.
If we consider an assembly of spheres in the continuum, the short-ranged
interaction between them only starts to become important when the neighborhood
of a considered particle is crowded.
In a lattice, this concept is usually quantified by counting the number of nearest neighbors of a particle on a lattice site.
Therefore, the threshold in the number of neighbors distinguishes
between crowded and not-crowded environments.
The softness provided by this Hamiltonian, then, in principle allows a
particle to be completely surrounded.
Thus, the model represents soft particles that can be highly packed with an energy cost.
This cost is proportional to the  repulsive potential and is higher
the lower the temperature.
The bidimensional sketch in Fig.~\ref{fig:meaning_softbm} shows a particle in  position $A$ which has some overlap with its neighbors.
The presence of such close neighbors implies that the particle in position $A$ has high energy and every movement to a position without overlaps (for example position $B$ in the figure) is favourable.
Thus, the kinetics of the system is characterized by a tendency of
particles to go from more to less crowded locations.

\subsection{The Kinetic Constraint}
It is known that nonergodic systems are characterized by a peculiar
behavior of particle movements.
Particles move inside a very small space for a typical characteristic time.
The reason is that the surrounding particles provide a cage within which the
movement of the considered particle is constrained.
Breaking of a cage requires a long and therefore improbable sequence of movements.
Occasionaly, the particle may escape from the cage and make longer movements before being trapped in another cage.
These two different behaviors are identifiable in  experiments
\cite{weeks2000} and reproducible by continuum simulations \cite{auer2003}.

\begin{figure}[htb]
\begin{minipage}[b]{0.49\columnwidth}
    \centering
    \subfigure[]{%
        \label{fig:subfig:caging}
        \includegraphics[width=0.8\columnwidth,angle=0]
        {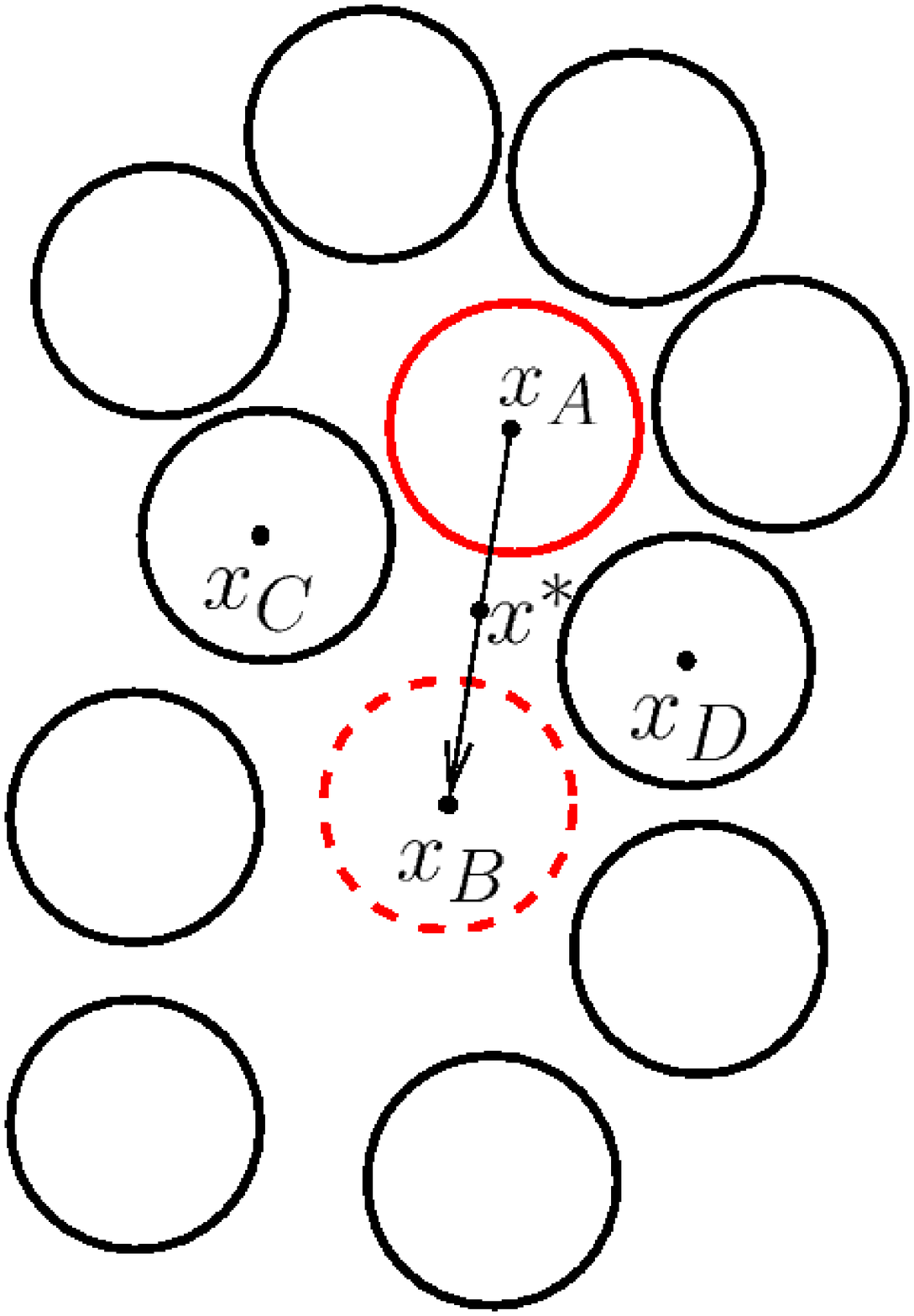}%
    }%
\end{minipage}%
\begin{minipage}[b]{0.49\columnwidth}
    \centering
    \subfigure[]{%
        \label{fig:subfig:barrier}
        \includegraphics[width=\columnwidth,angle=0]
        {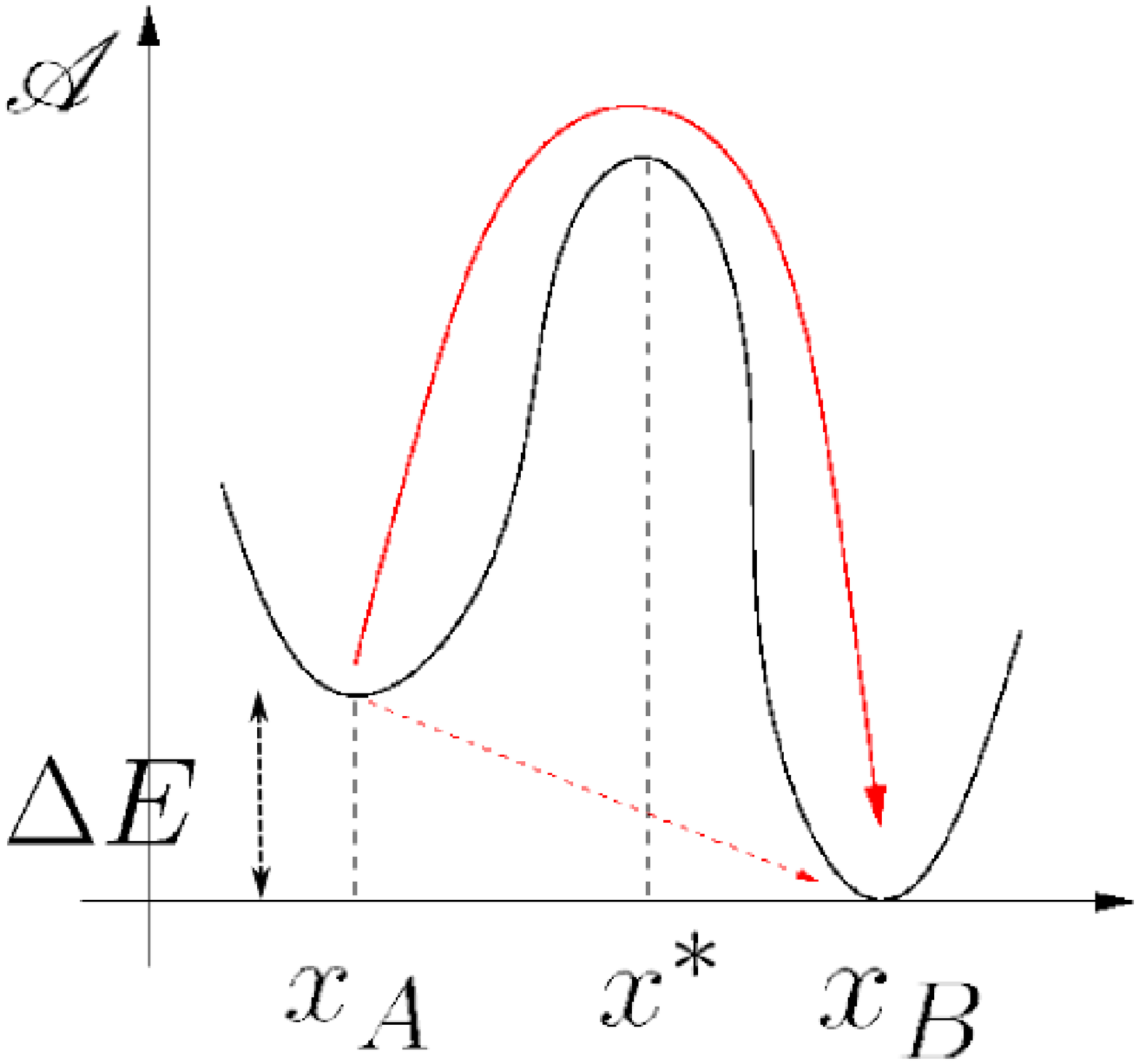}%
    }%
\end{minipage}
\caption{
(a) Schematic representation of the caging phenomenon. 
		To go from $x_A$ to $x_B$ the particle has to overcome the barrier (b) due to particles in the immediate vicinity (such as $C$ and $D$).
    Cage escape rates, that generally depend on the type of interaction and particle density, are represented by kinetic rates in Kob-Andersen models \cite{kob1993}.
    $\Delta E$ is the difference in local free energy between cages.
}
\label{fig:ka_meaning}
\end{figure}

The {\ka} model \cite{kob1993} correctly reproduces this intra-cage behavior of glassy systems, as well as many other important signatures such as blocked nonergodic states and dynamical heterogeneities \cite{pan2005,lawlor2005,degregorio2005}.
Being a lattice gas at the equilibrium, this model does not present any
ordered state or thermodinamic phase transition.
Our aim, then, is to implement a similar kinetic rule in our model and study the
interplay arising between arrest and crystallization.
Therefore, we define a kinetic constraint as follows.
% As in the {\ka } model, we define a parameter $c_K$.
A particle can only move from a site $i$ to a nearest neighboring site $j$ if the following rules are satisfied at the same time:
\begin{itemize}
\item[(a)] site $j$ is empty;
\item[(b)] the sum of the nearest and next-nearest neighbors of the particle in $i$ is at most equal to a fixed parameter $c_K$;
\item[(c)]  if the particle moves to $j$, the sum of its new nearest and next-nearest neighbors is at most equal to $c_K$.
\end{itemize}
This definition is based on the idea of associating the rate of motion from the central to a neighboring empty site with the number of particles that surround or cage it.
Thus we consider that caging by a  dense set of particles in the immediate surroundings of the central particle will last
longer (because an opening will present itself less frequently), and there is likely to be a threshold beneath which there are so
few particles that there is no effective caging at all.
As an example, the sketch in Fig.~\ref{fig:subfig:caging} represents a particle which is caged by several neighbors in such a way that the illustrated movement is only possible if the original cage of particles fluctuates and opens sufficiently to provide an exit path.
The relative unlikelihood of this originates in the fact that there are relatively few such configurations.
The local free energy will therefore reflect this by having a barrier between these two adjacent local minima (illustrated in Fig.~\ref{fig:subfig:barrier}).
By definition, then, this model aims to represent the $\alpha$-relaxation of glass-forming systems but does not describe $\beta$-relaxation.

The choice to also consider the next-nearest neighbors in the definition of the kinetic rule, unlike the {\ka} model, deserves a comment.
This choice is based on which particles on the lattice are caging.
If we consider the picture of the caging mechanism in Fig.~\ref{fig:subfig:caging}, the number of particles constituting the local cage is
larger than the number of particles that control the amount of space available for local (intra-cage) motion of the central particle.
The reason is that the typical size of a cage is even smaller than the radius of
the particles \cite{weeks2000}, involving only the closest neighbors, whereas the barrier can affect a larger number of particles, since a sequential chain of movements might be necessary in order to break a cage.
It is therefore natural that, in applying the kinetic constraint, we count the nearest- and next-nearest neighbors of the caged particle.
This definition is also interesting for a more technical reason.
Because of the discrete nature of the model, a wider range of values for $c_K$ allows a more satisfactory fine tuning.
The constraint can also be made soft, giving a finite probability for unfavourable moves.

% {\color{red}
This model, therefore, merges characteristics of both lattice gas models, such as the {\bm}, and kinetic models, as the {\ka}.
A recent paper presents a new {\bm}-type model with a three species mixture \cite{darst2010}. 
The composition is specifically designed to avoid crystallization and exhibits properties of a fragile glass-forming liquid.
The model successfully reproduces a stretched exponential law of the time relaxation, dynamical heterogeneities, Stokes-Einstein violation and other signatures.
% However, Darst et al.~do not distinguish the dynamics of the different types of particles and do not investigate the relationship between the dynamical heterogeneities and spatial arrangements of particle types.
The focus of this paper is to explore a model which explicitly includes ingredients motivated by current physical intuitions about glass-forming liquids: namely a short-ranged repulsion which determines an excluded volume and a kinetic rule which is a direct model of the cage effect.
Moreover, our model contains a genuine equilibrium crystal phase and allows a quite realistic comparison with colloidal systems which crystallize in the presence of low polydispersity \cite{pusey1986}.
Indeed, an investigation of the interplay between dynamical arrest and crystallization will be presented in a following publication \cite{cellai2011}.
% }

\subsection{Simulation Rules}

The two characteristics of the model can be implemented together in a Monte Carlo scheme in the following way.

The master equation for a dynamical process on a lattice reads
\begin{equation}
    \partial_t P(A,t) =
    \sum_B \left [ W_{B\to A} P(B,t) - W_{A\to B}P(A,t)\right],
    \label{eq:masterEquation}
\end{equation}
where $P(A,t)$ is the probability that the system is in the state $A$  at time $t$ \cite{frenkel2002book}.
In our case the states $A$ and $B$ are simply sets of occupancy numbers.
$W_{A\to B}$ is the rate of transitions from $A$ to $B$.
At the equilibrium, the left side of (\ref{eq:masterEquation}) is zero, then a condition that satisfies the master equation is that for every $B$:
\begin{equation}
    \frac{P_A^{eq}}{P_B^{eq}} = \frac{W_{A\to B}}{W_{B\to A}} =
                    \mathrm{e}^{-\beta(E_A - E_B)},
\end{equation}
where $E_A$, $E_B$ are the energies of the states $A$ and $B$, respectively.
The general expression for $W_{A\to B}$ is the product of a kinetic term $K$ and an energy term $F$:
\begin{equation}
  W_{A\to B} = K(A,B) F_{A\to B},
  \label{eq:wab}
\end{equation}
where
\begin{equation}
	\label{eq:fab}
  F_{A\to B} =
  \left\{
  \begin{array}{ll}
   \mathrm{e}^{-\beta(E_B-E_A)}  & (E_B-E_A)> 0\\
     1 & (E_B-E_A) < 0\\
  \end{array}
  \right. ,
\end{equation}
and $K(A,B)$ is a symmetric function with respect to $A$ and $B$.
It is evident that the choice of the function $K(A,B)$ does not affect the equilibrium state, but only the zones of the phase space visited during the simulation.
We only consider movements involving one particle at a time, with one lattice step displacement.
Thus, the only difference between state $A$ and $B$ consists of a single particle movement to an adjacent site.
Without any loss of generality, we can label the states $A$ and $B$ with the site indexes $i$ and $j$ of the places involved in the movement.
Hence, we rewrite the transition rate probability for a particle going from site $i$ to site $j$ as:
\begin{equation}
W_{i\to j} = K(i, j) F_{i\to j}.
\end{equation}

In a quite general way, we define the kinetic term as:
\begin{eqnarray}
K(i, j) & = & \exp\{-V_K [(m_i-c_K)\theta(m_i-c_K) +\nonumber\\
 & & +(m_j-c_K)\theta(m_j-c_K)]\}.
\label{eq:soft_kinetic_rule}
\end{eqnarray}
Here $m_i$ and $m_j$ are the sums of nearest and next-nearest neighbors of the particle before and after the movement, respectively, $\theta(x)$ is the Heaviside
function, $V_K$ is the barrier height (See Figure \ref{fig:ka_meaning}).
However, the results presented in this work deal with the hard limit of the constraint, which can be explicitly defined as:
\begin{equation}
K(i, j) = \theta(c_K-m_i)\theta(c_K-m_j),
\label{eq:kinetic_rule}
\end{equation}
where we assume the convention $\theta(0)\equiv 0$.
The soft kinetic rule (\ref{eq:soft_kinetic_rule}) simply provides a smoother behavior than the hard rule (\ref{eq:kinetic_rule}).
Results from simulations show that the soft rule may be used to interpolate between the discrete values of $c_K$, but it gives no additional interest to the model.

The  procedure described here aims to represent faithfully the kinetics of the model.
The  Monte Carlo scheme chosen is not supposed to reach  the equilibrium state quickly and efficiently (in fact, it is very slow).
On the contrary, our purpose is to represent the movements as they actually occur in a real system.
The main assumption that we make is that a single particle scheme is able to accomplish this task, the concern being the fact that real dynamics involves simultaneous movements of particles.
Indeed, it is known that dynamics can be realistically represented by physical-move kinetic Monte Carlo simulations when the  degrees of freedom considered are a slow subset of all of the degrees of freedom \cite{binder1997}.
This is correct in our case, especially because we consider high density systems, where the intra-cage movements are rare events.
In lattice models, a number of lattice sweeps is able to rebuild the average dynamics of the system, so that the macroscopic outcome is the same.

\section{Phase diagram and glass transition}

\begin{figure}
    \includegraphics[height=0.95\columnwidth,angle=270]{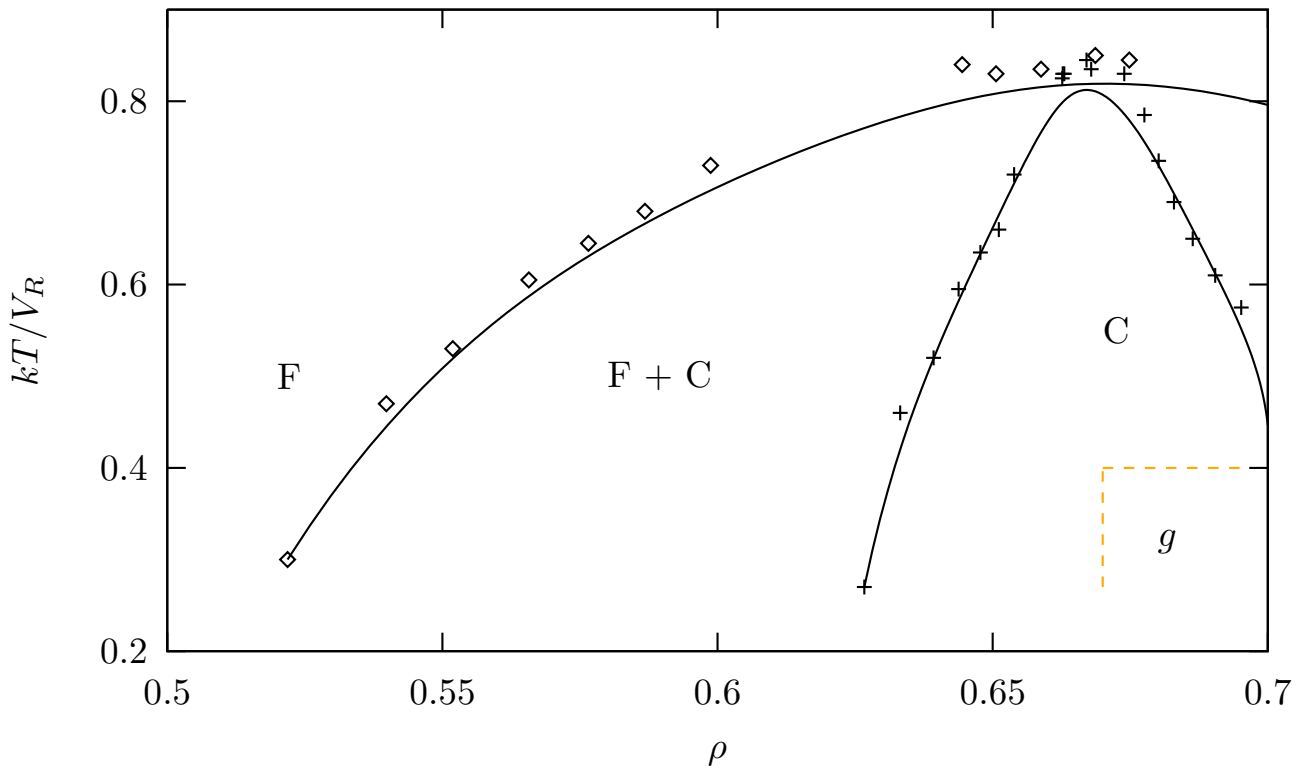}
\caption{
		Section of the equilibrium phase diagram of the model for
    $c_R=3$ \cite{mccullagh2005}, $F$=fluid, $C$=crystal.
    In the crystal phase for $T\leq0.4$ a glassy state is observed (for $c_K=10$). }
\label{fig:phasediag}
\end{figure}

% {\color{red}
The equilibrium phase diagram is not affected by the kinetic rules discussed above and has been already studied in  detail \cite{mccullagh2005}.
Here we give a summary of its properties.
For $c_R=3$, in a cubic lattice, the phase diagram is shown in Fig.\ \ref{fig:phasediag}.
The crystal phase is characterized by double layered diagonal planes with a periodicity of $\sqrt{3}$ lattice steps.
In the region involving a density roughly between 0.5 and 0.7 we observe the typical sequence of fluid, coexistence between fluid and crystal, and crystal.
At higher density the crystal re-melts into a fluid through a first-order phase transition which is driven by the high number of defects in the crystalline structure, making the disordered high-density fluid more entropically favoured \cite{mccullagh2005}.
% }

In this paper we will focus on $c_R=3$ and $c_K=10$.
These values are chosen in order to examine the most interesting cases for our purposes. For other parameter values, the model contains a rich variety of phenomena that we do not address here.

\begin{figure}[hbtp]
 \begin{flushright}
  \begin{overpic}[height=\columnwidth,angle=270]
   {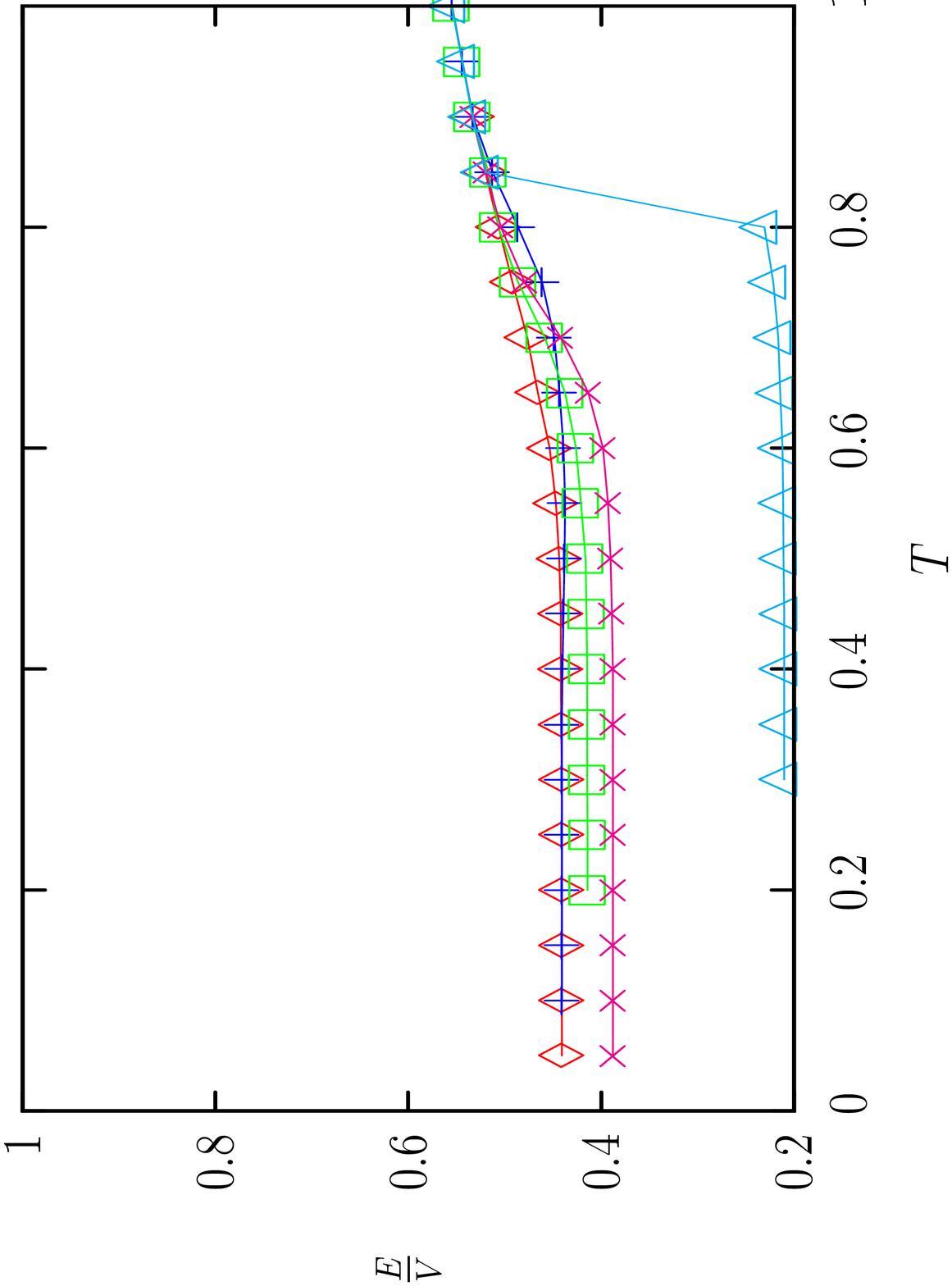}%
   \put(20,65){
            \includegraphics[width=0.35\columnwidth,angle=270]
                    {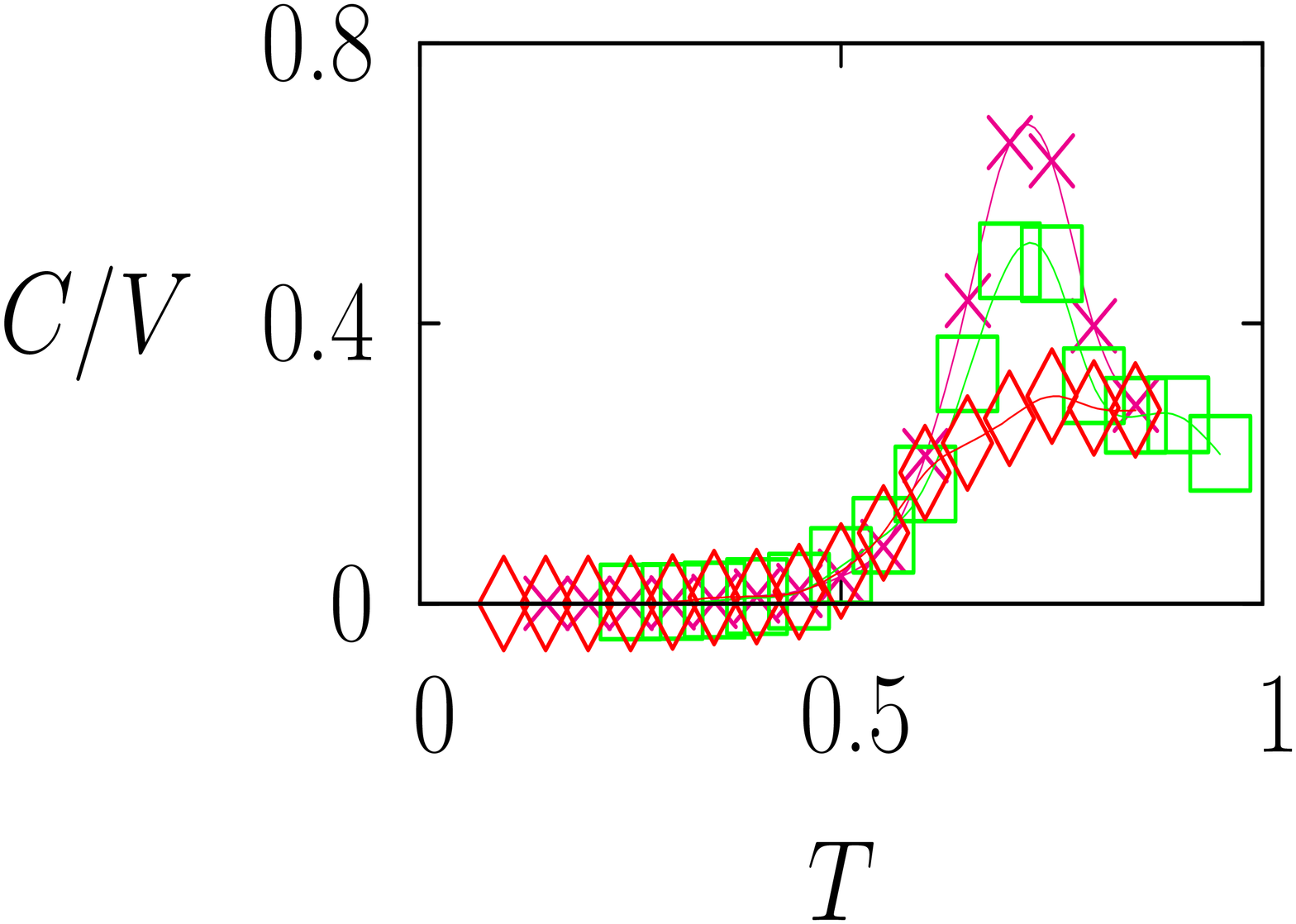}}
  \end{overpic}
    \caption{
    	Energy per particle with decreasing temperature.
			Monte Carlo simulations of a cubic lattice of size $30^3$ with  $c_K=10$ and $\rho=0.69$; lines are a guide for the eye.
			Time is measured in Monte Carlo Sweeps (MCS).
			For different cooling rates $R=\Delta T / \Delta t$
			($2.5\cdot10^{-7}$ ({$\Diamond$}), 
			$1.25\cdot10^{-7}$ ({$\Box$}), $5\cdot10^{-8}$ ({$\times$}) ),
			we observe behavior typical  of glass-forming liquids (see text).
			No arrest is present for $c_K=18$ ($R=5\cdot10^{-7}$, {$\triangle$}) 
			and the system quickly crystallizes.
			In one case ($R=2.5\cdot10^{-7}$, {$+$}) the temperature 
			has been increased from the arrested state.
			Hysteresis is observed.
			The inset shows the heat capacities for decreasing temperature
    	at three of the studied rates.	}
    \label{fig:rate_quenching_energies}
 \end{flushright}
\end{figure}

To illustrate effects commonly seen in real glasses, we have simulated the behavior of the temperature dependence of the mean
energy, for different cooling rates, at constant density.
For convenience's sake, we rescale the temperature to an adimensional variable $T = kT'/V_R$, and refer to this adimensional temperature in the rest of the paper.
As shown in Fig.\ \ref{fig:rate_quenching_energies}, for $\rho=0.69$ and $T\gtrsim
0.9$ (in the fluid phase according to Fig.\ \ref{fig:phasediag}), the system is only slightly slowed.
Between $T\simeq0.9$ and $T\simeq0.6$, just after crossing the liquid-crystal
first order transition, the energies decrease progressively and the system behaves as an under-cooled liquid. Below
$T\simeq0.6$, the energy is constant and far from the crystal energy. Decreasing the cooling rate, the picture remains much the same, 
though the energy plateau becomes progressively lower, as expected.
The states reached at $T\lesssim0.4$ consist of particles that are almost completely blocked, and using these states as
initial conditions in simulations even after long times ($t\sim10^6$ MCS) the configurations remain unchanged.

% {\color{red}
In Fig.~\ref{fig:rate_quenching_energies} we also show hysteresis between a cooling and a heating curve (at rate $R=2.5 10^{-7}$). In a cooling process the relaxation time increases until we reach a certain point where it becomes longer than the cooling step. This causes the system to fall out of equilibrium and it freezes into an arrested state at low temperature. As the direction of the structural relaxation process is always towards the equilibrium, when the temperature is increased, the $E-T$ curve follows a different path from the cooling curve, joining the equilibrium liquid curve after a small delay.
% }

The heat capacities shown in the inset of Fig.\ \ref{fig:rate_quenching_energies} are calculated from the derivatives of the energy plot. A small
quench-rate dependent peak is observed.

\section{Dynamical steady state behavior}

We  study the dynamics of the model on a cubic lattice, using the van Hove self-correlation function, i.e. the probability that a particle has traveled distance $r$ in time $t$.
In a lattice, the definition of $G_s$ is:
\begin{equation}
G_s(r,t) = \frac{1}{N} \sum_i^N \langle\delta_{\arrowvert \mathbf{r}_i(t)-\mathbf{r}_i(0)\arrowvert,r}
\rangle ,
\end{equation}
where $\delta$ is the Kronecker operator.

\begin{figure}[!htp]
	\begin{center}
		\includegraphics[height=\columnwidth,angle=-90]
		{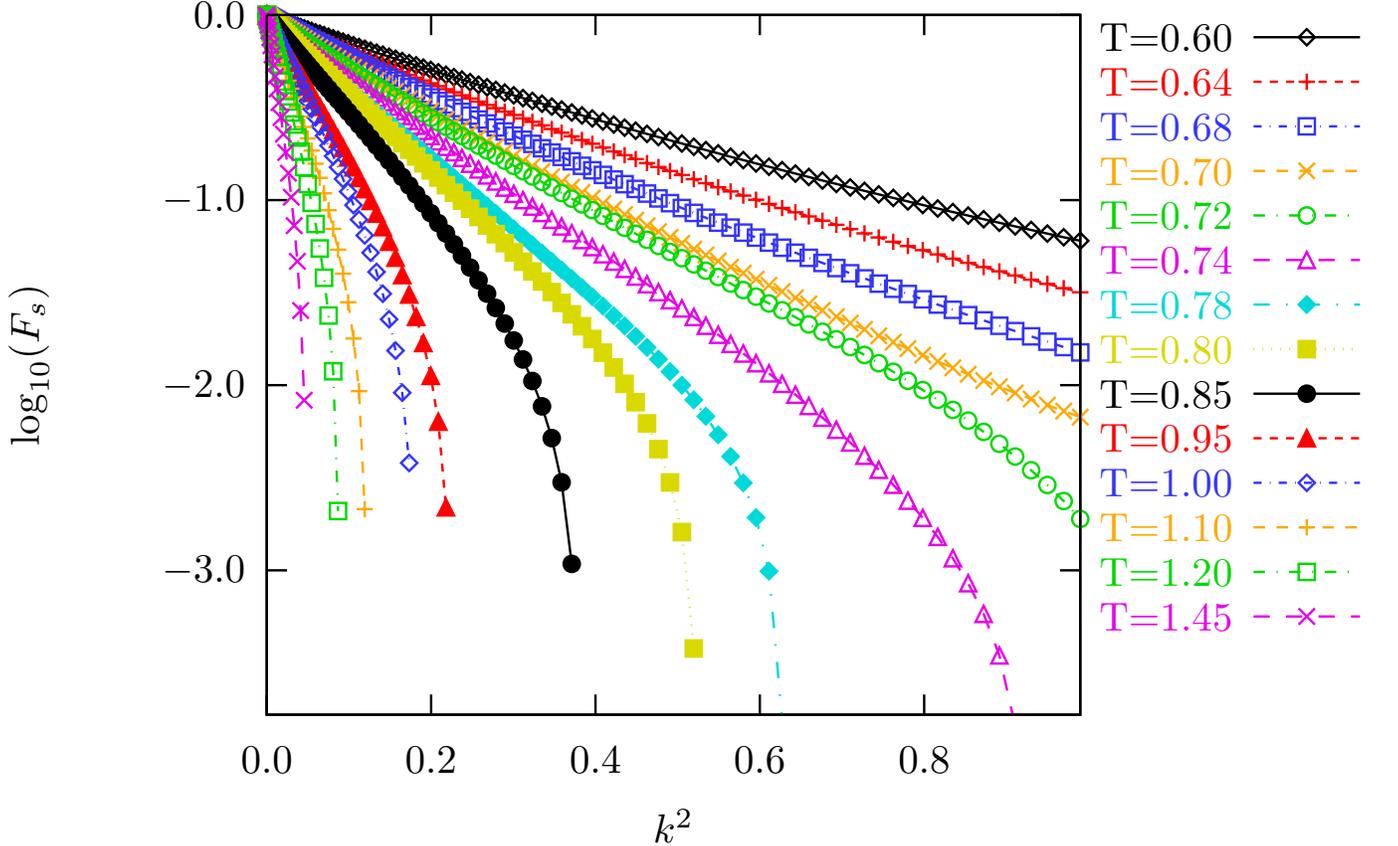}%
		\caption{
			Analysis of the $k$ dependence of $F_s(k,t)$ for $\rho=0.69$; lines are a guide for the eye.
			An equilibrated fluid at $T=1.5$ is used as initial state of simulations running at many temperatures from the range $T=0.60-1.45$. All the data refer to fixed time $t=200000$.
		}
	\label{fig:k_dependence_Fs}
	\end{center}
\end{figure}

\begin{figure}[!htp]
	\begin{center}
		\includegraphics[height=\columnwidth,angle=-90]
		{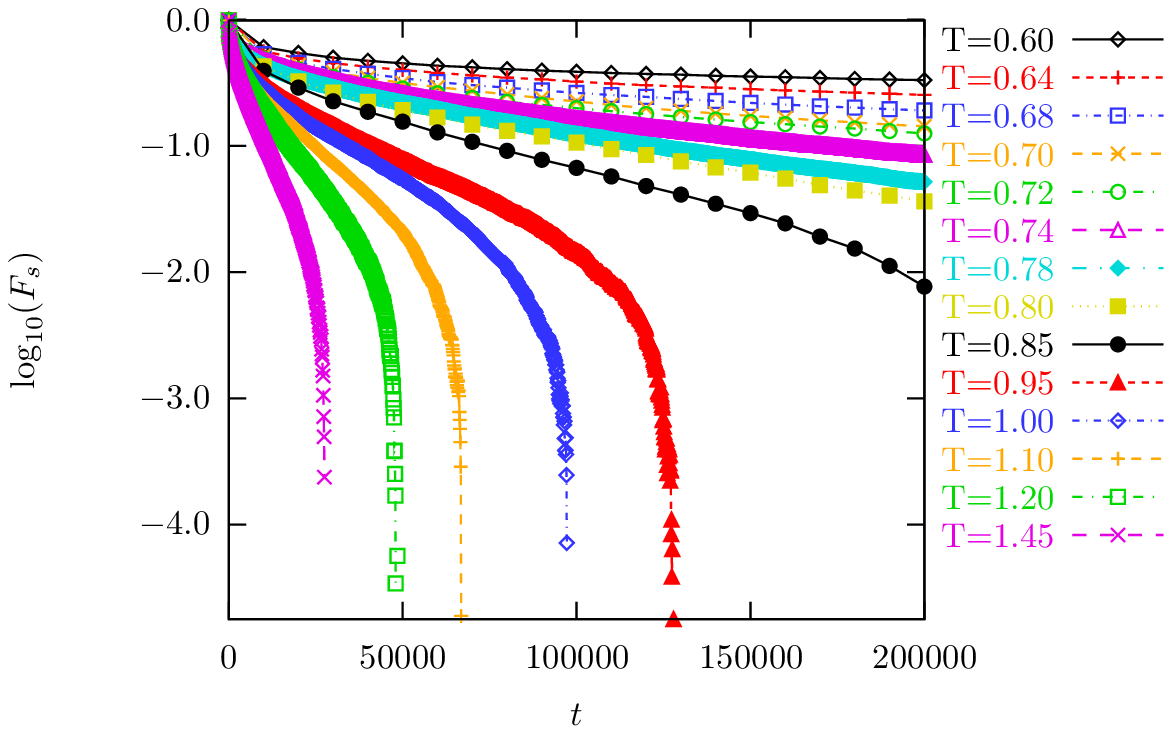}%
		\caption{
			Analysis of the time ($t$) dependence of $F_s(k,t)$ for fixed $k=0.5774$.
			As in Fig.~\ref{fig:k_dependence_Fs}, temperatures are in the range $T=0.60-1.45$.
		}
	\label{fig:t_dependence_Fs_k058}
	\end{center}
\end{figure}

\begin{figure}[!htp]
	\begin{center}
		\includegraphics[height=\columnwidth,angle=-90]
		{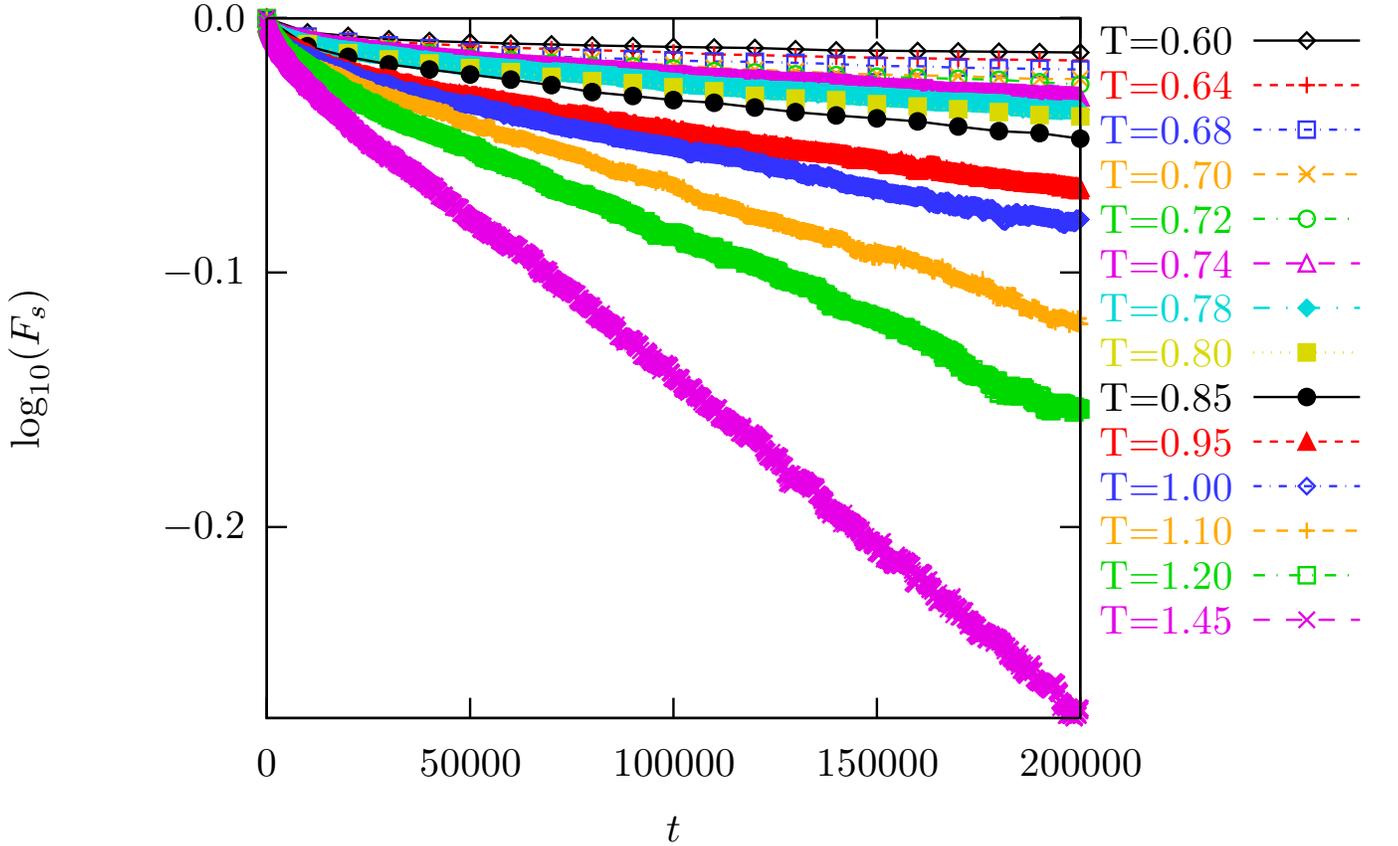}%
		\caption{
			Analysis of the time ($t$) dependence of $F_s(k,t)$ for fixed $k=0.0914$.
			As in Fig.~\ref{fig:k_dependence_Fs}, temperatures are in the range $T=0.60-1.45$.
		}
	\label{fig:t_dependence_Fs_k009}
	\end{center}
\end{figure}

The spatial Fourier transform $F_s(k,t)$ of $G_s(r,t)$ is named self intermediate scattering function and contitutes a useful quantity to study the relaxation properties of a system.
The onset of dynamical slowing is classically associated with $F_s(k,t)$ being
well fitted by the Kohlrausch-Williams-Watts (KWW) stretched exponential:
\begin{equation}
 F_s(k,t) =
A\;\mathrm{exp}\left[-\left(\frac{t}{\tau(T)}\right)^{\beta(T)}\right].
\label{eq:kww_law_fit}
\end{equation}
As one approaches the glass transition, typically $\tau$ diverges and $\beta$ decreases \cite{angell2000}.
We calculate $F_s$ and,
following a typical experimental approach, we fit the temperature dependence of the self intermediate scattering function by the KWW law.
It is known that for $k\to 0$, $F_s$ satisfies the gaussian approximation
\begin{equation}
		F_s(\k,t)=\exp\left\{ -\frac{1}{2d}k^2\langle \Delta\r(t)^2\rangle \right\},
		\label{eq:gaussian_approx_Fs}
\end{equation}
because after moving a  very large distance, particle movements always become uncorrelated \cite{binder2005}.
On the other hand, for $k\to\infty$, i.e. at very short distances, movements are always correlated, even in a perfectly diffusing liquid.
In order to study the glass transition, then, we have to choose an intermediate value of $k$ for which it is easy to check the validity of the KWW law (\ref{eq:kww_law_fit}), over the whole temperature range of interest.
Of course, in a finite lattice model the values of $k$ are both bounded and discrete, so that $k$ can be neither too small nor too large.
However, some values at small $k$ can still be affected by the gaussian approximation.
In order to check that, we consider a fluid state at high temperature ($T=1.5$) and then we run  simulations at the same density but at the temperatures studied in Fig.~\ref{fig:rate_quenching_energies}.
In Fig.~\ref{fig:k_dependence_Fs} plots of $F_s(k,t)$ versus $k^2$ are presented for fixed time $t=2\cdot 10^5$.
In our plots the label $k$ is  the wavenumber scaled in units of $2\pi$, so that $1/k$ directly corresponds to a displacement.
The linear initial behavior at small $k$ is compatible with the expectations.
At high $T$, the relaxation is obviously quicker, so that the crossover to the high $k$ regime is recognizable.
The fact that at low $T$ the behavior is gaussian at every $k$ does not imply that the system is diffusive, because,
as we will see further, the mean square displacement is not linear in time.
Figs.~\ref{fig:t_dependence_Fs_k058} and \ref{fig:t_dependence_Fs_k009}   represent the behavior of $F_s(k,t)$ versus time at fixed $k=0.577\simeq 1/\sqrt{3}$ (corresponding to the crystal periodicity) and $k\simeq 0.091$ (corresponding to 11 lattice spaces), respectively.
From Fig.~\ref{fig:t_dependence_Fs_k058} it transpires that at high temperatures the value of $F_s$ at $t=2\cdot10^5$ becomes extremely small, so that the observed behavior is affected by numerical error.
On the contrary, at $k=0.091$ (Fig.~\ref{fig:t_dependence_Fs_k009}), $F_s$ is finite at all temperatures and therefore allows an easy check of the stretched exponential law (\ref{eq:kww_law_fit}).

\begin{figure}[!htp]
	\begin{center}
		\includegraphics[height=0.95\columnwidth,angle=-90]
		{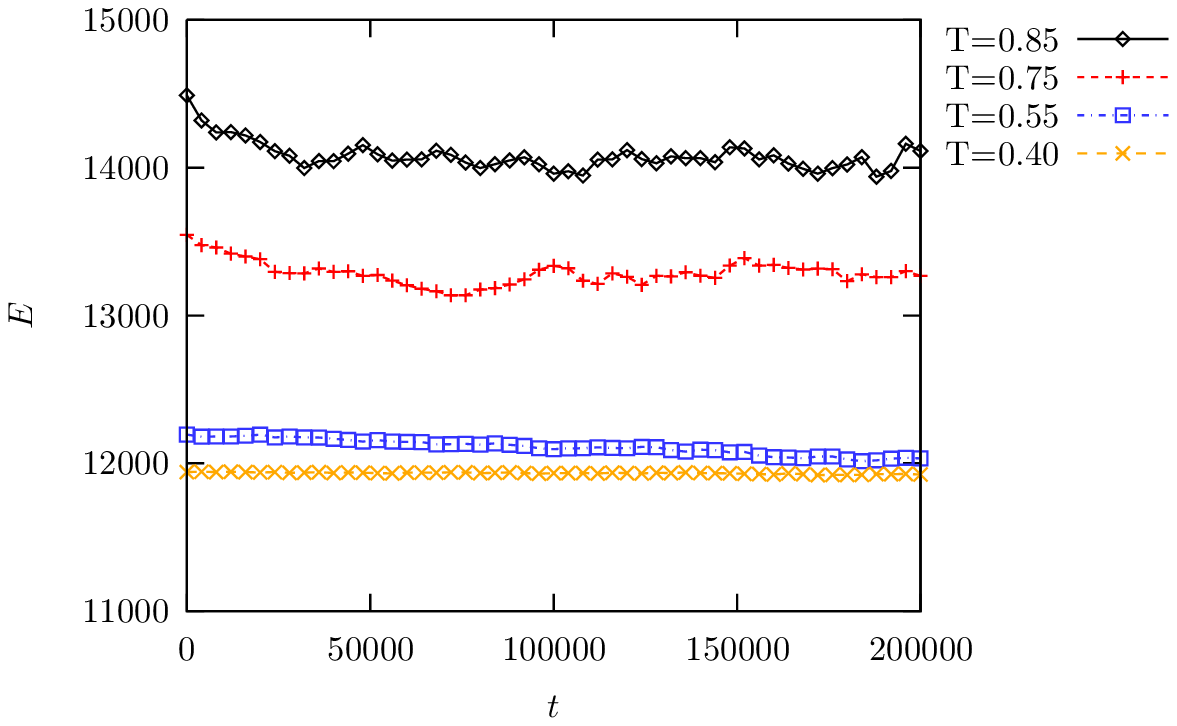}%
		\caption{
		Energy versus time for cooling rate R = $2.5\cdot10^{-6}$. Data from Fig.~\ref{fig:rate_quenching_energies}.
		}
	\label{fig:energy_time_rate_quenching}
	\end{center}
\end{figure}

The dynamical analysis we carry out is based on the assumption that the states involved are stationary.
This is an approximation because the system is actually aging slowly.
We check the validity of this assumption by plotting the energy evolution with time in the case of cooling at a fixed rate  $R = 2.5\cdot 10^{-6}$, for a few temperatures (Fig.~\ref{fig:energy_time_rate_quenching}).
% {\color{red}
It can be seen that the system is stationary for most temperatures with only a small amount of relaxation observed for $T=0.55$.
% }
We will examine the aging properties of the model in the following section.

\begin{figure}[!htp]
	\begin{center}
		\includegraphics[height=0.95\columnwidth,angle=270]
		{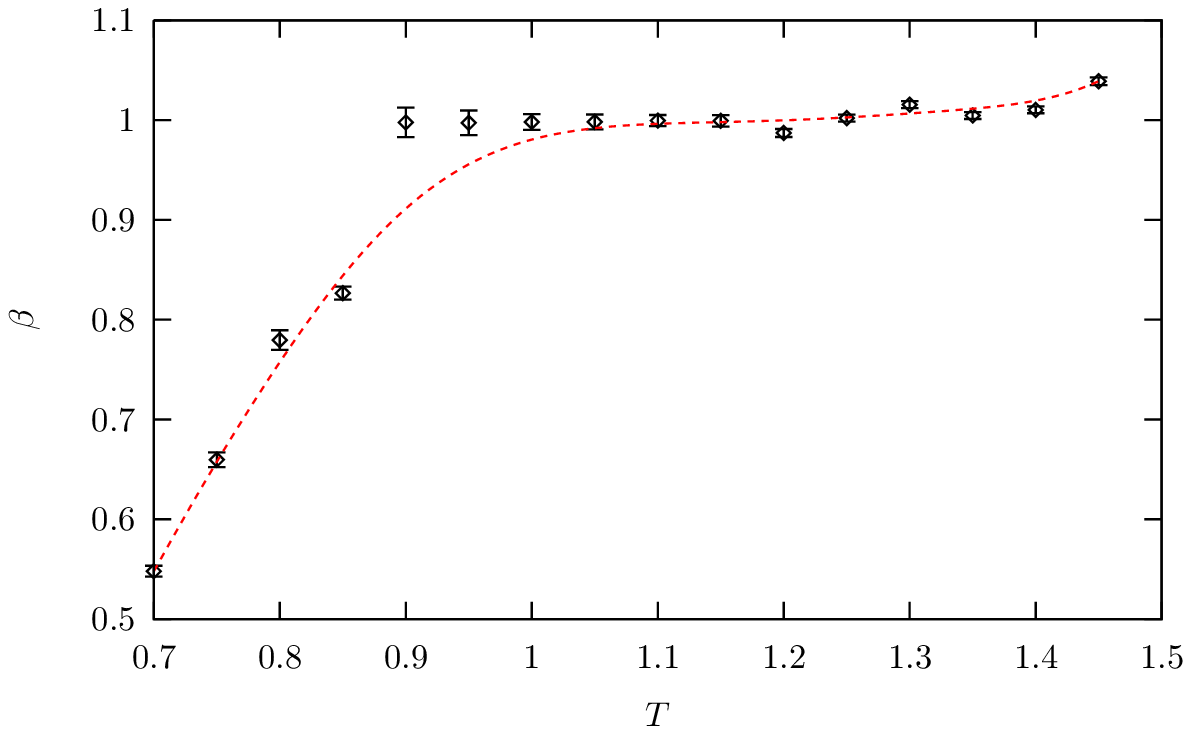}%
		\caption{
		Stretched exponent $\beta$ from the fitting of the KWW law (for $F_s(k,t)$) against $T$, for cooling simulations at the rate $R = 2.5\cdot 10^{-6}$; $\rho=0.69$ and $k=0.0914$.
		For high $T$ the decay is simply exponential ($\beta\simeq1$), whereas at $T\approx 1$ we observe a progressive deviation from this behavior.
		}
	\label{fig:T_beta}
	\end{center}
\end{figure}

\begin{figure}[!htp]
    \begin{center}
        \includegraphics[width=0.95\columnwidth,angle=0]
        {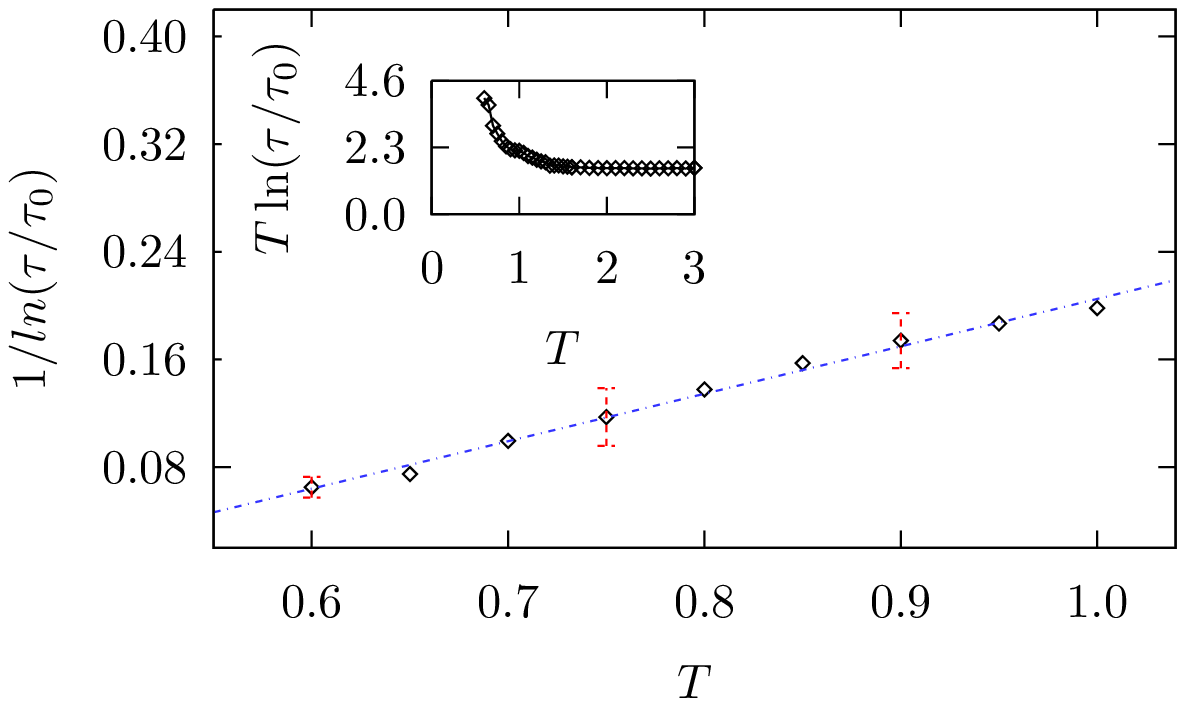}%
        \caption{Kauzmann plot for $k=0.0914$. 
        The Kauzmann temperature $T_K=0.42\pm0.03$ is obtained by extrapolation. Representative errors are marked on the graph, taken from tens of independent runs.
   Inset: Arrhenius plot (see e.g.\ \cite{sastry1998}).
    The quantity plotted, {$T\mathrm{ln}(\tau/\tau_0)$}, is constant when $\tau$ displays Arrhenius behavior.
    The infinite temperature relaxation time $\tau_0$ is obtained by fitting
    $\tau(T)$ values for $T=2.0-3.0$ to an Arrhenius form.
     Deviation from the constant,
    high-temperature value is seen around $T\approx 1$.
        }
    \label{fig:arrhenius}
    \end{center}
\end{figure}

In Fig.~\ref{fig:T_beta} the dependence of $\beta(T)$ on temperature is shown for $\rho=0.69$.
The set of data refers to a simulation of cooling performed taking as initial state the final state of the previous temperature.
We find that significant non-exponential slowing arises around $T=1.0$ and, consistent with expectations, the system develops non-Arrhenius behavior thereafter (see also inset to Fig.\ \ref{fig:arrhenius}) \cite{angell2000,sastry1998}.
This outcome is interesting, because it illustrates the fact that dynamical slowing can develop long before a true glass transition.

In order to estimate the Kauzmann temperature of the model, we  assume a classical dependence of the characteristic time on
temperature \cite{debenedetti1996}:
\begin{equation}
\tau=\tau_0\exp\left(\frac{A}{T-T_K}\right).
\label{eq:relaxation_VFT_law}
\end{equation}
The infinite temperature relaxation time $\tau_0$ is given by fitting the values $\tau(T)$ at $2.0<T<3.0$ to an Arrhenius form \cite{sastry1998}.
In Fig.~\ref{fig:arrhenius} the fit of the law in (\ref{eq:relaxation_VFT_law}) is calculated for $\rho=0.69$.
The extrapolation of the data leads to a divergence  at $T_K=0.42\pm0.03$ which would then be considered the glass
transition.

\begin{figure}[!htp]
	\begin{center}
	\includegraphics[height=0.95\columnwidth,angle=270]
	{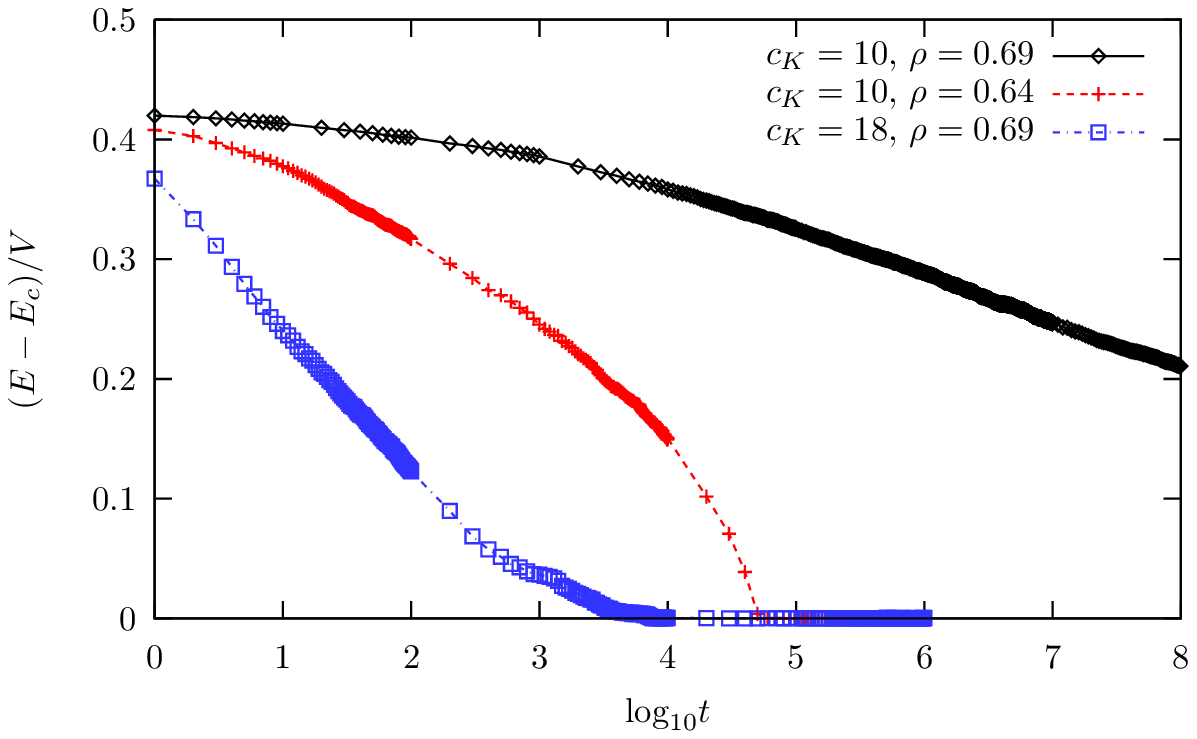}%
	\caption{Energy evolution of the model with time ($c_K=10$, size $30^3$).
	Energy difference with respect to the crystal energy is plotted for densities $\rho=0.64$ and $\rho=0.69$, at $T=0.4$.
	Starting from a fluid state, the time scale is dramatically different in the two cases: for $\rho=0.64$, the system crystallizes in less than $10^5$ MCS, for $\rho=0.69$, an extremely slow evolution is observed.
	Setting $c_K=18$ for $\rho=0.69$, i.e.\ removing the kinetic constraint, the system very quickly crystallizes.
	}
	\label{fig:energies}
	\end{center}
\end{figure}

% Bassler story
% {\color{red}
The extrapolated value of the Kauzmann temperature $T_K$ corresponds in our model to a temperature where disordered configurations remain disordered for a time scale much longer than the one of the fast processes.
Only slow aging remains, as we are going to show further (Fig.~\ref{fig:energies}), and the dynamics is sub-diffusive with no tendency to change (as shown by Fig.~\ref{fig:msd069}).
This is compatible with a scenario of strong space correlations due to caging and vanishing configurational entropy.

\begin{figure}[htb]
\begin{minipage}[b]{0.49\columnwidth}
    \centering
    \subfigure[]{%
        \label{fig:subfig:lowT}
        \includegraphics[height=\columnwidth,angle=270]{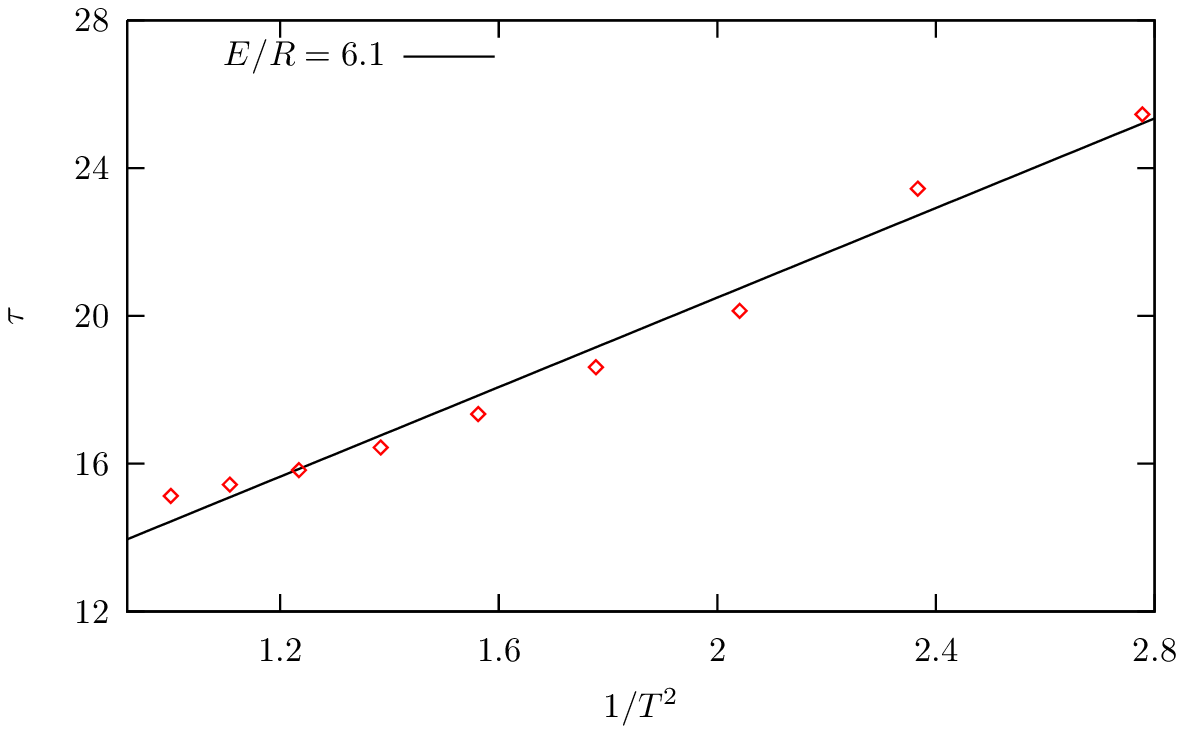}
    }%
\end{minipage}%
\begin{minipage}[b]{0.49\columnwidth}
    \centering
    \subfigure[]{%
        \label{fig:subfig:highT}
        \includegraphics[height=\columnwidth,angle=270]{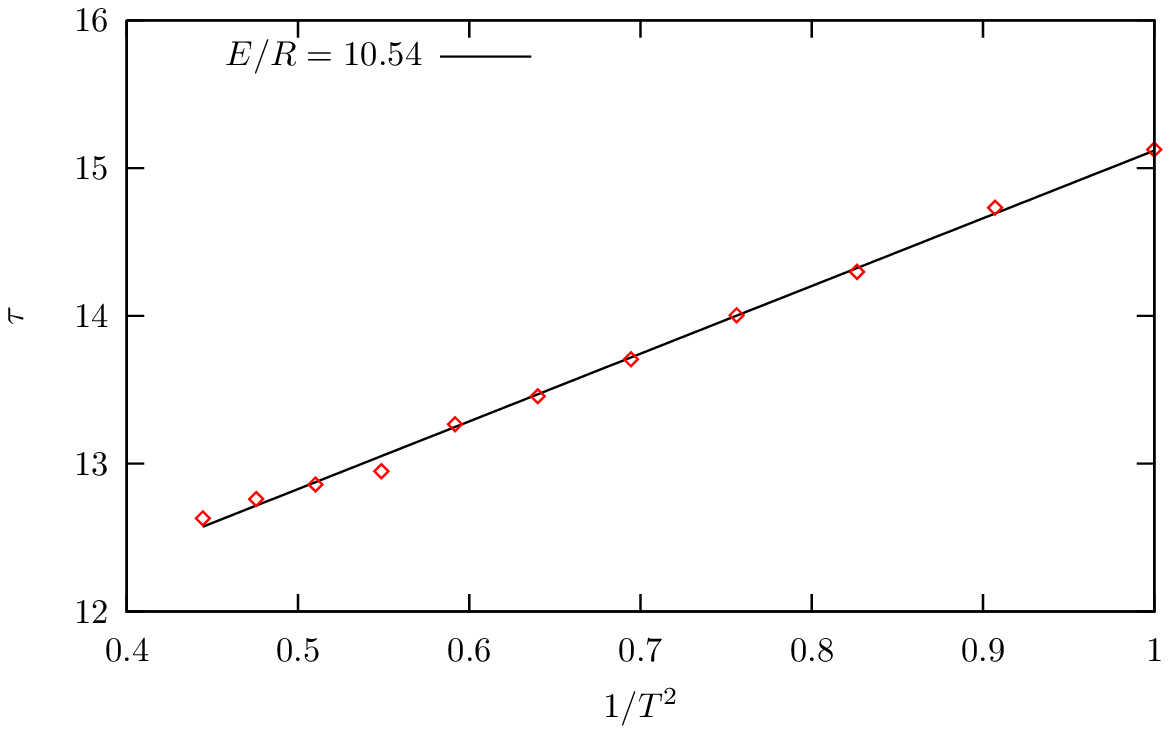}
    }%
\end{minipage}
\caption{B\"assler law fitted for low (a) and high (b) temperatures.
}
\label{fig:bassler}
\end{figure}

Other functional forms can fit the data with some approximation.
In Fig.~\ref{fig:bassler} we plot our data in the B\"assler form $\tau = A\exp(E/RT^2)$ \cite{bassler1987}: the fit is quite good at high $T$, but it slightly worsens in approaching the supercooled region, as the dynamics slows down and it moves away from a normal fluid.
% 
% \begin{figure}[htb]
%     \centering
%     \subfloat[]{%
%         \label{fig:subfig:lowT}
%         \includegraphics[height=0.95\columnwidth,angle=270]{lowT-tau-bassler}
%     }%
%     \subfloat[]{%
%         \label{fig:subfig:highT}
%         \includegraphics[height=0.95\columnwidth,angle=270]{highT-tau-bassler}
%     }%
% \caption{B\"assler law fitted for low (a) and high (b) temperatures.
% }
% \label{fig:bassler}
% \end{figure}
% 

% }

Let us focus now on the mobility of the system in the glassy region of the phase diagram.
We consider a fluid state at high temperature ($T=1.5$) and then we run  simulations at the same density but at much lower temperature (for example $T=0.4$), where the crystal is the equilibrium state and at the onset of the hypothetical glass transition (Fig.~\ref{fig:phasediag}).
Fig.~\ref{fig:energies} shows the evolution of energies for two representative densities, on the left and on the right hand side of the hypothetical arrest transition.

It is important to realize that, though only $c_R$ and $V_R$ are relevant to the equilibrium phase behavior, all three parameters, including $c_K$, arise from underlying microscopic interactions, and a general description of maxima, saddles and minima of a potential surface is impossible without them.
Thus, in the absence of the caging kinetic rules ($c_K=18$) the system quite quickly crystallizes (Fig.~\ref{fig:energies}).
Furthermore, for relatively low densities (e.g.~$\rho=0.64$) the system crystallizes almost immediately because caging is ineffective.
On the other hand, for $\rho=0.69$ and at low temperature the system evolves so slowly that crystallization has not been observed in any accessible time for our simulations ($t \sim 10^{8}$).
To our knowledge, this dramatic glass-like freezing of the system is present in lattice models only when both short-ranged repulsion and barrier crossing effects are combined as in the present model.

% IS
% {\color{red}
The kinetic rule is a fundamental part of the model, as the Hamiltonian in itself does not incorporate the caging phenomenon and therefore the dynamical heterogeneities typical of quasi arrested systems.
We believe that this kinetic rule is well justified by experiments which show the presence of caging and dynamical heterogeneities \cite{weeks2000,kegel2000}, and our model should be considered part of this tradition.
Moreover, there is extensive evidence that these types of kinetically constrained models reproduce a remarkable set of dynamical heterogeneities \cite{lawlor2005,toninelli2005}.

\begin{figure}[hbt]
	\begin{center}
		\includegraphics[height=0.95\columnwidth,angle=270]{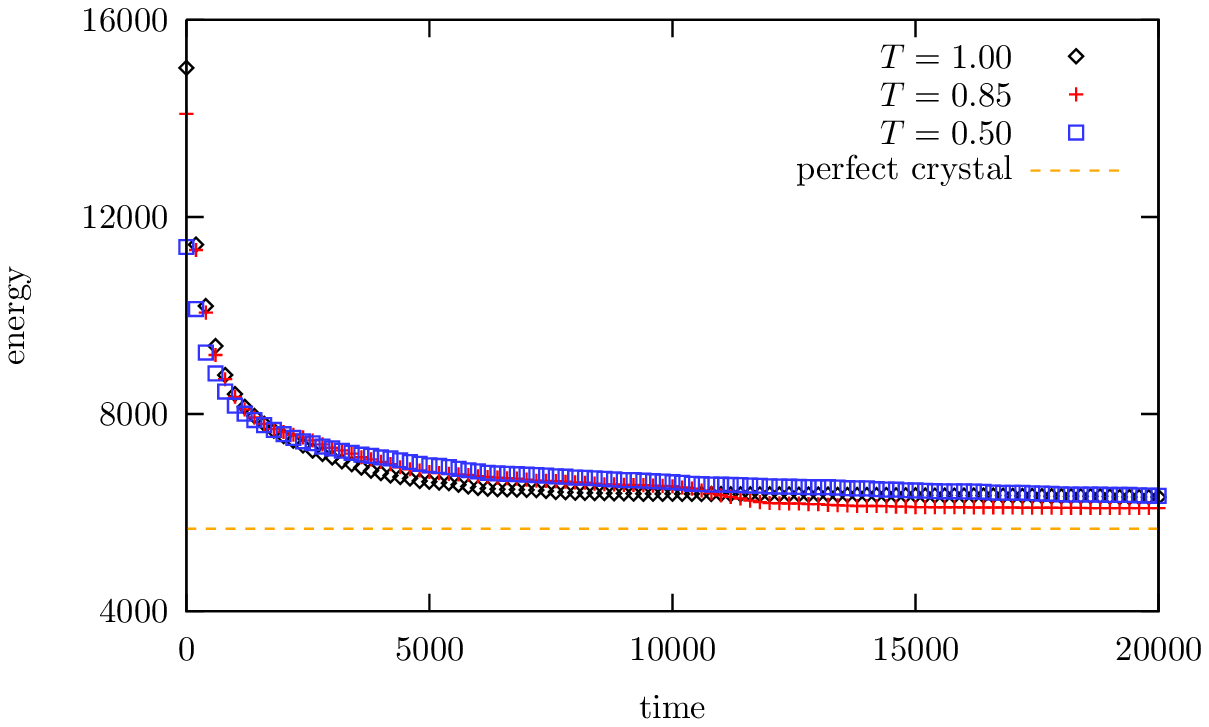}%
	\end{center}
\caption{Inherent structure energy relaxation of samples at different temperatures $T$.}
\label{fig:is-relaxation}
\end{figure}

To better capture the onset of dynamical slowing, some authors perform an inherent structure (IS) analysis which has been successful in both continuum and lattice models \cite{sastry1998, glotzer2000b}.
However, this method appears to be suitable only for models which are fully described by a Hamiltonian.
Following the IS implementation for lattice models illustrated in \cite{glotzer2000b}, we consider an algorithm which is going through every particle in typewriter order and try in sequence a movement in all the six possible directions:
\begin{itemize}
	\item If $\Delta E < 0$ and the kinetic rule allows it, do the move.
	\item If $\Delta E=0$ and the kinetic rule allows it, do the move with probability $1/2$.
	\item In any other case, reject the move.
\end{itemize}
However, this procedure leads to a crystalline state in our model, as is shown by the IS calculation of samples at different temperatures in Fig.~\ref{fig:is-relaxation}.
The reason is that in our model the interplay between short-ranged repulsion and dynamical heterogeneities is partially destroyed by this analysis, because the systematic search of lower energy movements promotes unlikely paths, and destroys long-lived islands of quasi-blocked particles.
In other words, the systematic search of the IS analysis allows us to find the very few movements which unlock the whole system but which would be very unlikely in a stochastic approach.
% }

\section{Aging}

\begin{figure}[htbp]
 \begin{center}
	\includegraphics[height=0.95\columnwidth,angle=270]{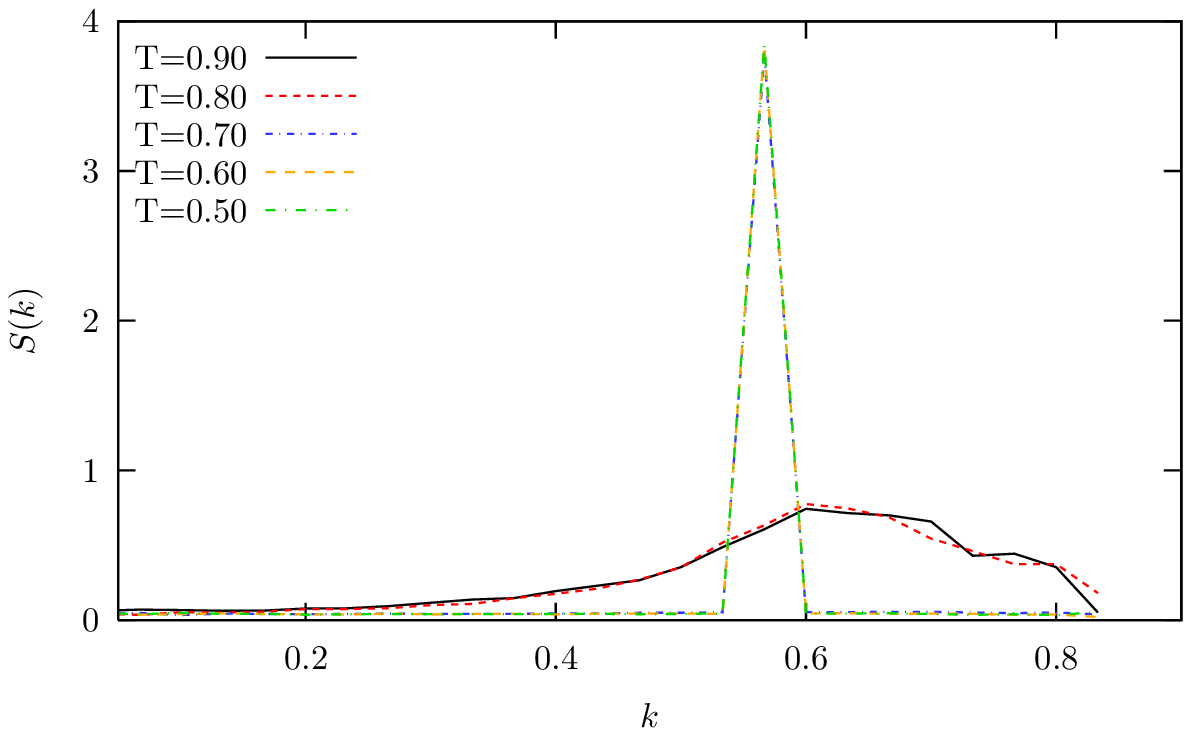}
	\caption{
	Static structure factor versus time for a cooling simulation at $\rho=0.64$ and rate $R=2.5\cdot10^6$.
	}
 \label{fig:Sk_rho064}
 \end{center}
\end{figure}

\begin{figure}[htbp]
 \begin{center}
	\includegraphics[height=0.95\columnwidth,angle=270]{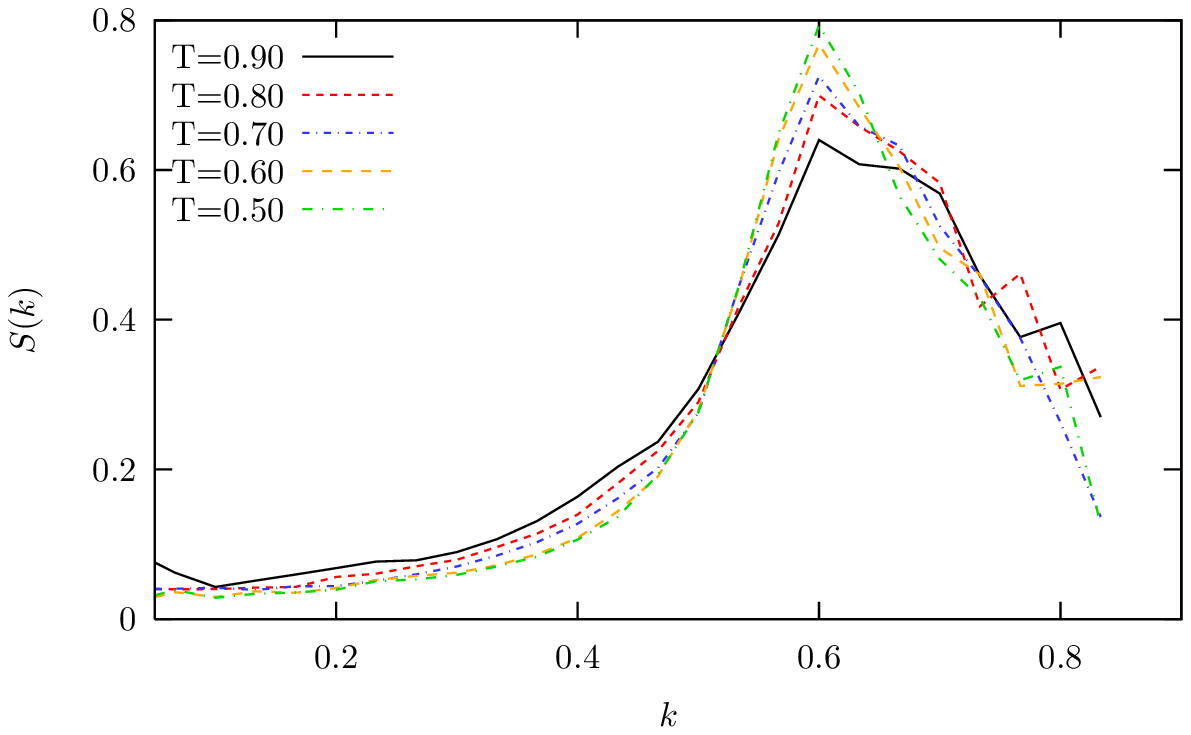}
  \caption{
	Static structure factor versus time for the cooling simulation at $\rho=0.69$ and rate $R=2.5\cdot10^6$ as in Fig.~\ref{fig:rate_quenching_energies}.
	}
 \label{fig:Sk_rho069}
 \end{center}
\end{figure}

The time evolution of the energy as in Fig.~\ref{fig:energies} clearly shows that the system ages, even at very high density.
Indeed, we know that the model presents a crystal phase at the equilibrium and therefore aging phenomena are expected.
We have already seen that for high $c_K$ the system crystallizes.
However, at $c_K=10$, we do not observe crystallization at high density.
In spite of that, the system can be heterogeneous, containing little crystallites in an amorphous background.
In order to understand better this issue, we study the static structure factor $S(k)$.
In Figures \ref{fig:Sk_rho064} and \ref{fig:Sk_rho069} we show the evolution of the structure factor during the process of cooling.
At low density ($\rho=0.64$), the system quickly develops a single peak corresponding to the crystal periodicity of $\sqrt{3}$ lattice spacings.
At high density ($\rho=0.69$), the structure factor does not present the previous typical behavior.
The profile of $S(k)$ is much lower, as expected from direct observation of the configurations, but, on the other hand, it is evident that some degree of order is present.

\begin{figure}[htbp]
 \begin{center}
	\includegraphics[height=0.95\columnwidth,angle=270]{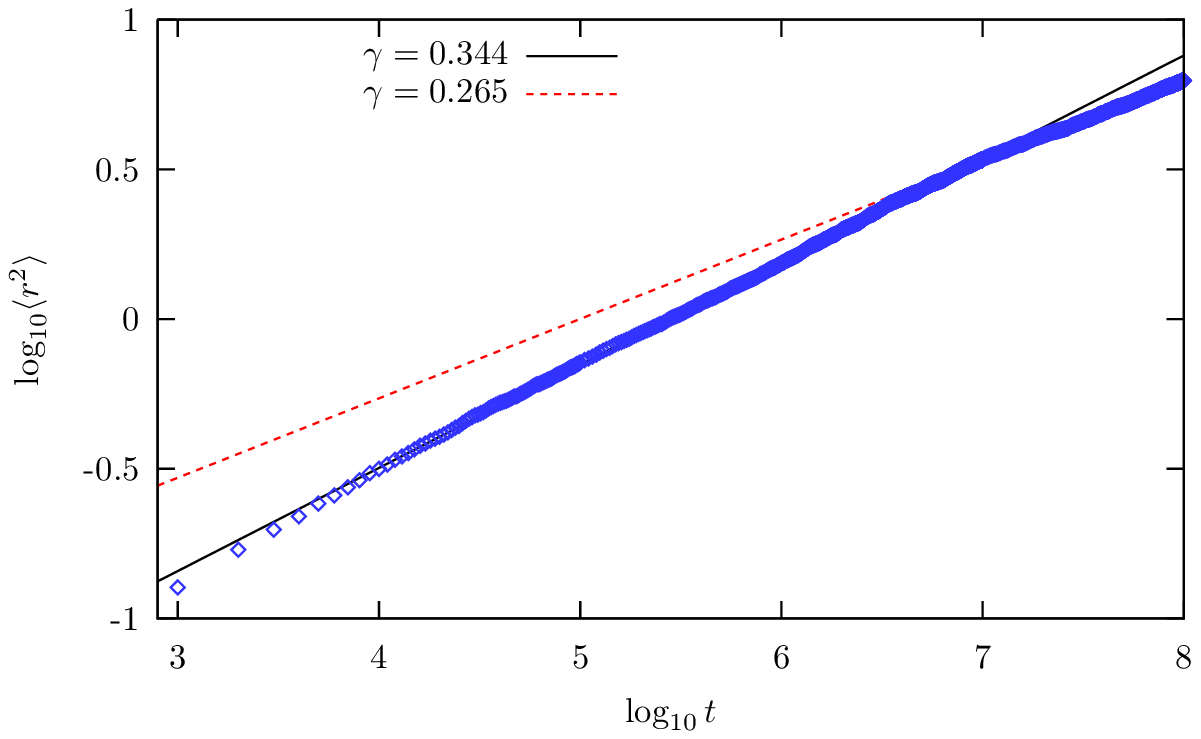}
    \caption{
    Mean squared displacement for $\rho=0.69$ of simulations at fixed temperature $T=0.4$ starting from an equilibrated fluid at higher temperature ($T=1.5$).
    There appear to be two power law behaviors with a crossover at $t\sim 10^7$.
    For earlier times the exponent is 0.344, later it becomes 0.265.
	}
 \label{fig:msd069}
 \end{center}
\end{figure}

In Fig.~\ref{fig:msd069}, the mean square displacement is plotted versus time for a system at $\rho=0.69$.
Up to $t\sim 10^7$, the data are well fitted by a power law $\langle r^2\rangle \sim t^{\gamma}$, with $\gamma=0.344$.
This value is remarkably similar to the one found in experiments involving different types of particles \cite{simeonova2004,yu2009}.
The fact that the agreement is also quantitatively acceptable could be the signature of a new universality class in the area of arrested matter.
To our knowledge, this is a new achievement for a lattice model.
The second regime presents a value $\gamma=0.265$ which is related to aging of the sample.
Comparing this plot with the energy evolution in Fig.~\ref{fig:energies}, it appears that the breakdown of the subdiffusive law $t^{0.34}$ is connected with a change of convexity in the energy evolution at about $t\approx 10^7$.
This regime not only shows that the system slows down with time, but also indicates that it evolves to a disordered, perhaps completely blocked, non-crystalline state.
It is also possible to note that for higher densities the characteristic time separating the two power  laws shifts to lower values.
Therefore, the $\gamma=0.34$ law seems to be characteristic of the arrest transition, because moving beyond the arrest transition restricts its domain of validity.

We can interpret the slowing of the system beyond arrest as caused by the existence of a new kind of state dynamically similar to random close packing.
Particle movements do not lead to configurations that are more and more similar to the crystalline equilibrium state.
On the contrary, the pathway makes the system evolve into a sort of side-track: the slowing process is progressive, perhaps leading to a completely arrested state.

Our interpretation of a subdiffusive exponent close to $1/3$, based on the analysis above, is that the phenomenon happens only when the free energy landscape has some particular characteristics.
First, it has to present a global minimum corresponding to a stable crystal phase.
As a consequence, the temperature has to be low enough to guarantee quasi-hard particle interaction.
Second, there must be many local minima, separated by high barriers, so that the time evolution to the global minimum requires long times to overcome or circumvent the barriers.
Therefore, it is clear that, in order to observe the phenomenon, the parameters of the model have to be tuned in such a way that these conditions are satisfied.
It is still questionable how the value of the subdiffusive exponent (which could be a rational number) can be theoretically justified.
In this paper, we only show that this law, observed in a few different experiments, can even be reproduced by a simple lattice model, pointing out that the phenomenon seems to be a robust property of the arrest transition.

As a last remark, from Figure \ref{fig:msd069} it emerges that the mean square displacement can be relatively high.
After $10^8$ time steps, we have $\sqrt{\langle r^2\rangle} \approx 2.2$, which means that on average every particle has traveled more than 2 lattice steps.
In spite of setting $T\approx T_K $, this should not be surprising, because a glassy behavior is perfectly compatible with subdiffusivity.
The fact that the diffusion constant is zero does not mean that every particle is blocked, but only that there is not a characteristic time in the distribution of wait times per particle \cite{terraneo2009}.

\section{Conclusions}
The model may be used to study many other interesting properties,
but discussion of the details are not central to our point here.
Rather we note that, for the first time, a simple lattice model has been able to reproduce the range of phenomena from real glasses and energy landscape models, including rate dependent slowing of the energy, divergence of the characteristic time with an appropriate KWW law, vanishing of the diffusion constant, onset of collective behavior, and a possible quantitative representation of subdiffusivity in dense systems.
It is interesting to explore the origins of these new effects in the model.

Models based on kinetic constraints alone create a complex effective free energy landscape in which, at sufficiently high density, many movements are prohibited by infinite or large barriers.
Nonetheless, many dynamical pathways involving
long ranged transport still remain at fixed (zero) energy.
True dynamical arrest only occurs in such models when the lattice
is fully filled. This dramatic reduction of dynamical pathways induced by kinetic constraints certainly leads to dynamical
slowing, but not to true glassy behavior. If we now allow different local energies within different cages one obtains a
complex energy landscape. Then, rare pathways that were formerly barrier-less remain `easy', but acquire a multitude of
smaller energy barriers. The accumulation of such bumps against the backdrop of a vanishing number of easy pathways ultimately
leads to interesting singular behavior for the characteristic time, which is considered to be truly representative of the
glassy state.
This is the root of glass behavior, and its physical origins are quite clear in our model.

Regarding the behavior of the mean square displacement, lattice models clearly show that subdiffusion arises when particle movements are intermittent and there is not a characteristic wait time \cite{terraneo2009}.
In a dense system, subsequent movements are allowed by the same hole, coming back according to a random walk scheme.
Therefore, arbitrary long return paths cause the absence of a  characteristic time.
This behavior changes when particles are being moved by several different holes, and we observe diffusion, or crystallization occurs.
However, this behavior does not break down at the arrest transition.
If the system has diverging subdiffusive domains, there is not a time scale for their breaking.
Moreover, the kinetic properties of our model are due to the underlining Kob-Andersen kinetic rules, which have been thoroughly studied in the past \cite{kob1993}.
In such kind of kinetic models, it is possible to identify a dynamical correlation length which describes the typical size of domains where local movements of particles do not contribute to diffusion \cite{degregorio2005}.
When this length diverges, the system becomes dynamically arrested.
The presence of a characteristic length in our model could give an explanation for the remarkable behavior of colloids under gravity, where the gravitational length does not capture the nature of the phenomenon \cite{simeonova2004}.

It must be considered intriguing that the ingredients for such a model are already known in the literature but that they have 
not hitherto been combined in this way.
Purely repulsive Hamiltonian lattice models without kinetical barriers 
\cite{biroli2002,ciamarra2003b,weigt2003}
appear not to yield a KWW characteristic time law, as in experiments and continuous simulations.
On the other hand, purely kinetic models do not have a crystal phase
\cite{kob1993}.
Here we have a simple model that
reproduces the main effects associated with glassy systems. Thereby it opens up the possibility of a more transparent dialogue
between experimental scientists, simulators in off-lattice models, and theorists working in such highly simplified models.

We acknowledge with pleasure discussions with W. Kegel and S. Granick.
This paper is written within an EU-US research consortium, funded in part by a grant from the Marie Curie program of the
European Union, (`Arrested Matter', contract number MRTN-CT-2003-504712), and by an SFI grant.

% \bibliography{biblio}

%Merlin.mbs v4.21 2009-07-09.
\begin{thebibliography}{}%
\makeatletter
\providecommand \@ifxundefined [1]{%
 \ifx #1\undefined \expandafter \@firstoftwo
 \else \expandafter \@secondoftwo
\fi
}%
\providecommand \@ifnum [1]{%
 \ifnum #1\expandafter \@firstoftwo
 \else \expandafter \@secondoftwo
\fi
}%
\providecommand \enquote [1]{``#1''}%
\providecommand \bibnamefont  [1]{#1}%
\providecommand \bibfnamefont [1]{#1}%
\providecommand \citenamefont [1]{#1}%
\providecommand\href[0]{\@sanitize\@href}%
\providecommand\@href[1]{\endgroup\@@startlink{#1}\endgroup\@@href}%
\providecommand\@@href[1]{#1\@@endlink}%
\providecommand \@sanitize [0]{\begingroup\catcode`\&12\catcode`\#12\relax}%
\@ifxundefined \pdfoutput {\@firstoftwo}{%
 \@ifnum{\z@=\pdfoutput}{\@firstoftwo}{\@secondoftwo}%
}{%
 \providecommand\@@startlink[1]{\leavevmode\special{html:<a href="#1">}}%
 \providecommand\@@endlink[0]{\special{html:</a>}}%
}{%
 \providecommand\@@startlink[1]{%
  \leavevmode
  \pdfstartlink
   attr{/Border[0 0 1 ]/H/I/C[0 1 1]}%
   user{/Subtype/Link/A<</Type/Action/S/URI/URI(#1)>>}%
  \relax
 }%
 \providecommand\@@endlink[0]{\pdfendlink}%
}%
\providecommand \url  [0]{\begingroup\@sanitize \@url }%
\providecommand \@url [1]{\endgroup\@href {#1}{\urlprefix}}%
\providecommand \urlprefix [0]{URL }%
\providecommand \Eprint[0]{\href }%
\@ifxundefined \urlstyle {%
  \providecommand \doi [1]{doi:\discretionary{}{}{}#1}%
}{%
  \providecommand \doi [0]{doi:\discretionary{}{}{}\begingroup
  \urlstyle{rm}\Url }%
}%
\providecommand \doibase [0]{http://dx.doi.org/}%
\providecommand \Doi[1]{\href{\doibase#1}}%
\providecommand \bibAnnote [3]{%
  \BibitemShut{#1}%
  \begin{quotation}\noindent
    \textsc{Key:}\ #2\\\textsc{Annotation:}\ #3%
  \end{quotation}%
}%
\providecommand \bibAnnoteFile [2]{%
  \IfFileExists{#2}{\bibAnnote {#1} {#2} {\input{#2}}}{}%
}%
\providecommand \typeout [0]{\immediate \write \m@ne }%
\providecommand \selectlanguage [0]{\@gobble}%
\providecommand \bibinfo [0]{\@secondoftwo}%
\providecommand \bibfield [0]{\@secondoftwo}%
\providecommand \translation [1]{[#1]}%
\providecommand \BibitemOpen[0]{}%
\providecommand \bibitemStop [0]{}%
\providecommand \bibitemNoStop [0]{.\EOS\space}%
\providecommand \EOS [0]{\spacefactor3000\relax}%
\providecommand \BibitemShut [1]{\csname bibitem#1\endcsname}%
%</preamble>
\end{thebibliography}%


\begin{thebibliography}{10}%
\makeatletter
\providecommand \@ifxundefined [1]{%
 \ifx #1\undefined \expandafter \@firstoftwo
 \else \expandafter \@secondoftwo
\fi
}%
\providecommand \@ifnum [1]{%
 \ifnum #1\expandafter \@firstoftwo
 \else \expandafter \@secondoftwo
\fi
}%
\providecommand \enquote [1]{``#1''}%
\providecommand \bibnamefont  [1]{#1}%
\providecommand \bibfnamefont [1]{#1}%
\providecommand \citenamefont [1]{#1}%
\providecommand\href[0]{\@sanitize\@href}%
\providecommand\@href[1]{\endgroup\@@startlink{#1}\endgroup\@@href}%
\providecommand\@@href[1]{#1\@@endlink}%
\providecommand \@sanitize [0]{\begingroup\catcode`\&12\catcode`\#12\relax}%
\@ifxundefined \pdfoutput {\@firstoftwo}{%
 \@ifnum{\z@=\pdfoutput}{\@firstoftwo}{\@secondoftwo}%
}{%
 \providecommand\@@startlink[1]{\leavevmode}%
 \providecommand\@@endlink[0]{}%
}{%
 \providecommand\@@startlink[1]{%
  \leavevmode
  \pdfstartlink
   attr{/Border[0 0 1 ]/H/I/C[0 1 1]}%
   user{/Subtype/Link/A<</Type/Action/S/URI/URI(#1)>>}%
  \relax
 }%
 \providecommand\@@endlink[0]{\pdfendlink}%
}%
\providecommand \url  [0]{\begingroup\@sanitize \@url }%
\providecommand \@url [1]{\endgroup\@href {#1}{\urlprefix}}%
\providecommand \urlprefix [0]{URL }%
\providecommand \Eprint[0]{\href }%
\@ifxundefined \urlstyle {%
  \providecommand \doi [1]{doi:\discretionary{}{}{}#1}%
}{%
  \providecommand \doi [0]{doi:\discretionary{}{}{}\begingroup
  \urlstyle{rm}\Url }%
}%
\providecommand \doibase [0]{http://dx.doi.org/}%
\providecommand \Doi[1]{\href{\doibase#1}}%
\providecommand \bibAnnote [3]{%
  \BibitemShut{#1}%
  \begin{quotation}\noindent
    \textsc{Key:}\ #2\\\textsc{Annotation:}\ #3%
  \end{quotation}%
}%
\providecommand \bibAnnoteFile [2]{%
  \IfFileExists{#2}{\bibAnnote {#1} {#2} {\input{#2}}}{}%
}%
\providecommand \typeout [0]{\immediate \write \m@ne }%
\providecommand \selectlanguage [0]{\@gobble}%
\providecommand \bibinfo [0]{\@secondoftwo}%
\providecommand \bibfield [0]{\@secondoftwo}%
\providecommand \translation [1]{[#1]}%
\providecommand \BibitemOpen[0]{}%
\providecommand \bibitemStop [0]{}%
\providecommand \bibitemNoStop [0]{.\EOS\space}%
\providecommand \EOS [0]{\spacefactor3000\relax}%
\providecommand \BibitemShut [1]{\csname bibitem#1\endcsname}%
%</preamble>
\bibitem{angell2000}%
  \BibitemOpen
  \bibfield{author}{%
  \bibinfo {author} {\bibfnamefont{C.~A.}\ \bibnamefont{Angell}}, \bibinfo
  {author} {\bibfnamefont{K.~L.}\ \bibnamefont{Ngai}}, \bibinfo {author}
  {\bibfnamefont{G.~B.}\ \bibnamefont{McKenna}}, \bibinfo {author}
  {\bibfnamefont{P.~F.}\ \bibnamefont{McMillan}},\ and\ \bibinfo {author}
  {\bibfnamefont{S.~W.}\ \bibnamefont{Martin}},\ }%
  \bibfield{journal}{%
  \bibinfo {journal} {J. Appl. Phys.}\ }%
  \textbf{\bibinfo {volume} {88}},\ \bibinfo {pages} {3113} (\bibinfo {year}
  {2000})%
  \bibAnnoteFile{NoStop}{angell2000}%
\bibitem{binder2005}%
  \BibitemOpen
  \bibfield{author}{%
  \bibinfo {author} {\bibfnamefont{K.}~\bibnamefont{Binder}}\ and\ \bibinfo
  {author} {\bibfnamefont{W.}~\bibnamefont{Kob}},\ }%
  \emph{\bibinfo {title} {Glassy Materials and disordered Solids}}\ (\bibinfo
  {publisher} {World Scientific, Singapore},\ \bibinfo {year} {2005})%
  \bibAnnoteFile{NoStop}{binder2005}%
\bibitem{kob1993}%
  \BibitemOpen
  \bibfield{author}{%
  \bibinfo {author} {\bibfnamefont{W.}~\bibnamefont{Kob}}\ and\ \bibinfo
  {author} {\bibfnamefont{H.~C.}\ \bibnamefont{Andersen}},\ }%
  \bibfield{journal}{%
  \bibinfo {journal} {Phys. Rev. E}\ }%
  \textbf{\bibinfo {volume} {48}},\ \bibinfo {pages} {4364} (\bibinfo {year}
  {1993})%
  \bibAnnoteFile{NoStop}{kob1993}%
\bibitem{pal2008}%
  \BibitemOpen
  \bibfield{author}{%
  \bibinfo {author} {\bibfnamefont{P.}~\bibnamefont{Pal}}, \bibinfo {author}
  {\bibfnamefont{C.~S.}\ \bibnamefont{O'Hern}}, \bibinfo {author}
  {\bibfnamefont{J.}~\bibnamefont{Blawzdziewicz}}, \bibinfo {author}
  {\bibfnamefont{E.~R.}\ \bibnamefont{Dufresne}},\ and\ \bibinfo {author}
  {\bibfnamefont{R.}~\bibnamefont{Stinchcombe}},\ }%
  \bibfield{journal}{%
  \Doi{10.1103/PhysRevE.78.011111}{\bibinfo {journal} {Physical Review E}}\ }%
  \textbf{\bibinfo {volume} {78}},\ \bibinfo {pages} {011111+} (\bibinfo
  {month} {Jul}\ \bibinfo {year} {2008}),\
  \url{http://dx.doi.org/10.1103/PhysRevE.78.011111}%
  \bibAnnoteFile{NoStop}{pal2008}%
\bibitem{jackle1994}%
  \BibitemOpen
  \bibfield{author}{%
  \bibinfo {author} {\bibfnamefont{J.}~\bibnamefont{J\"ackle}}\ and\ \bibinfo
  {author} {\bibfnamefont{A.}~\bibnamefont{Kr\"onig}},\ }%
  \bibfield{journal}{%
  \bibinfo {journal} {J. Phys.: Condens. Matter}\ }%
  \textbf{\bibinfo {volume} {6}},\ \bibinfo {pages} {7633} (\bibinfo {year}
  {1994})%
  \bibAnnoteFile{NoStop}{jackle1994}%
\bibitem{mezard2000}%
  \BibitemOpen
  \bibfield{author}{%
  \bibinfo {author} {\bibfnamefont{M.}~\bibnamefont{M\'ezard}}\ and\ \bibinfo
  {author} {\bibfnamefont{G.}~\bibnamefont{Parisi}},\ }%
  \bibfield{journal}{%
  \bibinfo {journal} {J. Phys. Condens. Matter}\ }%
  \textbf{\bibinfo {volume} {12}},\ \bibinfo {pages} {6655} (\bibinfo {year}
  {2000})%
  \bibAnnoteFile{NoStop}{mezard2000}%
\bibitem{toninelli2006}%
  \BibitemOpen
  \bibfield{author}{%
  \bibinfo {author} {\bibfnamefont{C.}~\bibnamefont{Toninelli}}, \bibinfo
  {author} {\bibfnamefont{G.}~\bibnamefont{Biroli}},\ and\ \bibinfo {author}
  {\bibfnamefont{D.}~\bibnamefont{Fisher}},\ }%
  \bibfield{journal}{%
  \bibinfo {journal} {Phys. Rev. Lett.}\ }%
  \textbf{\bibinfo {volume} {96}} (\bibinfo {year} {2006})%
  \bibAnnoteFile{NoStop}{toninelli2006}%
\bibitem{liu1998}%
  \BibitemOpen
  \bibfield{author}{%
  \bibinfo {author} {\bibfnamefont{A.~J.}\ \bibnamefont{Liu}}\ and\ \bibinfo
  {author} {\bibfnamefont{S.~R.}\ \bibnamefont{Nagel}},\ }%
  \bibfield{journal}{%
  \bibinfo {journal} {Nature}\ }%
  \textbf{\bibinfo {volume} {396}},\ \bibinfo {pages} {21} (\bibinfo {year}
  {1998})%
  \bibAnnoteFile{NoStop}{liu1998}%
\bibitem{sellitto2006}%
  \BibitemOpen
  \bibfield{author}{%
  \bibinfo {author} {\bibfnamefont{M.}~\bibnamefont{Sellitto}},\ }%
  \bibfield{journal}{%
  \bibinfo {journal} {Phys. Rev. B}\ }%
  \textbf{\bibinfo {volume} {73}},\ \bibinfo {pages} {180202} (\bibinfo {year}
  {2006})%
  \bibAnnoteFile{NoStop}{sellitto2006}%
\bibitem{mehta2000}%
  \BibitemOpen
  \bibfield{author}{%
  \bibinfo {author} {\bibfnamefont{A.}~\bibnamefont{Mehta}}\ and\ \bibinfo
  {author} {\bibfnamefont{G.~C.}\ \bibnamefont{Barker}},\ }%
  \bibfield{journal}{%
  \bibinfo {journal} {J. Phys. Condens. Matter}\ }%
  \textbf{\bibinfo {volume} {12}},\ \bibinfo {pages} {6619} (\bibinfo {year}
  {2000})%
  \bibAnnoteFile{NoStop}{mehta2000}%
\bibitem{biroli2002}%
  \BibitemOpen
  \bibfield{author}{%
  \bibinfo {author} {\bibfnamefont{G.}~\bibnamefont{Biroli}}\ and\ \bibinfo
  {author} {\bibfnamefont{M.}~\bibnamefont{M\'ezard}},\ }%
  \bibfield{journal}{%
  \bibinfo {journal} {Phys. Rev. Lett.}\ }%
  \textbf{\bibinfo {volume} {88}},\ \bibinfo {pages} {025501} (\bibinfo {year}
  {2002})%
  \bibAnnoteFile{NoStop}{biroli2002}%
\bibitem{bouchaud2003}%
  \BibitemOpen
  \bibfield{author}{%
  \bibinfo {author} {\bibfnamefont{R.~A.}\ \bibnamefont{Denny}}, \bibinfo
  {author} {\bibfnamefont{D.~R.}\ \bibnamefont{Reichman}},\ and\ \bibinfo
  {author} {\bibfnamefont{J.-P.}\ \bibnamefont{Bouchaud}},\ }%
  \bibfield{journal}{%
  \bibinfo {journal} {Phys. Rev. Lett.}\ }%
  \textbf{\bibinfo {volume} {90}},\ \bibinfo {pages} {025503} (\bibinfo {year}
  {2003})%
  \bibAnnoteFile{NoStop}{bouchaud2003}%
\bibitem{garrahan2002}%
  \BibitemOpen
  \bibfield{author}{%
  \bibinfo {author} {\bibfnamefont{J.~P.}\ \bibnamefont{Garrahan}}\ and\
  \bibinfo {author} {\bibfnamefont{D.}~\bibnamefont{Chandler}},\ }%
  \bibfield{journal}{%
  \bibinfo {journal} {Phys. Rev. Lett.}\ }%
  \textbf{\bibinfo {volume} {89}},\ \bibinfo {pages} {035704} (\bibinfo {year}
  {2002})%
  \bibAnnoteFile{NoStop}{garrahan2002}%
\bibitem{ritort2003}%
  \BibitemOpen
  \bibfield{author}{%
  \bibinfo {author} {\bibfnamefont{F.}~\bibnamefont{Ritort}}\ and\ \bibinfo
  {author} {\bibfnamefont{P.}~\bibnamefont{Sollich}},\ }%
  \bibfield{journal}{%
  \bibinfo {journal} {Adv. Phys.}\ }%
  \textbf{\bibinfo {volume} {52}},\ \bibinfo {pages} {219} (\bibinfo {year}
  {2003})%
  \bibAnnoteFile{NoStop}{ritort2003}%
\bibitem{starr2002}%
  \BibitemOpen
  \bibfield{author}{%
  \bibinfo {author} {\bibfnamefont{F.~W.}\ \bibnamefont{Starr}}, \bibinfo
  {author} {\bibfnamefont{S.}~\bibnamefont{Sastry}}, \bibinfo {author}
  {\bibfnamefont{J.~F.}\ \bibnamefont{Douglas}},\ and\ \bibinfo {author}
  {\bibfnamefont{S.~C.}\ \bibnamefont{Glotzer}},\ }%
  \bibfield{journal}{%
  \bibinfo {journal} {Phys. Rev. Lett.}\ }%
  \textbf{\bibinfo {volume} {89}},\ \bibinfo {pages} {125501} (\bibinfo {year}
  {2002})%
  \bibAnnoteFile{NoStop}{starr2002}%
\bibitem{gotze1991}%
  \BibitemOpen
  \bibfield{author}{%
  \bibinfo {author} {\bibfnamefont{W.}~\bibnamefont{G\"otze}},\ }%
  in\ \emph{\bibinfo {booktitle} {Freezing and Glass Transition}},\ \bibinfo
  {editor} {edited by\ \bibinfo {editor} {\bibfnamefont{J.~P.}\
  \bibnamefont{Hansen}}, \bibinfo {editor}
  {\bibfnamefont{D.}~\bibnamefont{Levesque}},\ and\ \bibinfo {editor}
  {\bibfnamefont{J.}~\bibnamefont{Zinn-Justin}}}\ (\bibinfo {publisher}
  {Amsterdam: North Holland},\ \bibinfo {year} {1991})\ p.\ \bibinfo {pages}
  {287}%
  \bibAnnoteFile{NoStop}{gotze1991}%
\bibitem{giovambattista2003}%
  \BibitemOpen
  \bibfield{author}{%
  \bibinfo {author} {\bibfnamefont{N.}~\bibnamefont{Giovambattista}}, \bibinfo
  {author} {\bibfnamefont{S.~V.}\ \bibnamefont{Buldyrev}}, \bibinfo {author}
  {\bibfnamefont{F.~W.}\ \bibnamefont{Starr}},\ and\ \bibinfo {author}
  {\bibfnamefont{H.~E.}\ \bibnamefont{Stanley}},\ }%
  \bibfield{journal}{%
  \bibinfo {journal} {Phys. Rev. Lett.}\ }%
  \textbf{\bibinfo {volume} {90}},\ \bibinfo {pages} {085506} (\bibinfo {year}
  {2003})%
  \bibAnnoteFile{NoStop}{giovambattista2003}%
\bibitem{lawlor2005}%
  \BibitemOpen
  \bibfield{author}{%
  \bibinfo {author} {\bibfnamefont{A.}~\bibnamefont{Lawlor}}, \bibinfo {author}
  {\bibfnamefont{P.}~\bibnamefont{{De Gregorio}}}, \bibinfo {author}
  {\bibfnamefont{P.}~\bibnamefont{Bradley}}, \bibinfo {author}
  {\bibfnamefont{M.}~\bibnamefont{Sellitto}},\ and\ \bibinfo {author}
  {\bibfnamefont{K.~A.}\ \bibnamefont{Dawson}},\ }%
  \bibfield{journal}{%
  \bibinfo {journal} {Phys. Rev. E}\ }%
  \textbf{\bibinfo {volume} {72}},\ \bibinfo {pages} {021401} (\bibinfo {year}
  {2005})%
  \bibAnnoteFile{NoStop}{lawlor2005}%
\bibitem{degregorio2005}%
  \BibitemOpen
  \bibfield{author}{%
  \bibinfo {author} {\bibfnamefont{P.}~\bibnamefont{{De Gregorio}}}, \bibinfo
  {author} {\bibfnamefont{A.}~\bibnamefont{Lawlor}}, \bibinfo {author}
  {\bibfnamefont{P.}~\bibnamefont{Bradley}},\ and\ \bibinfo {author}
  {\bibfnamefont{K.~A.}\ \bibnamefont{Dawson}},\ }%
  \bibfield{journal}{%
  \bibinfo {journal} {Proc. Nat. Acad. Sci.}\ }%
  \textbf{\bibinfo {volume} {102}},\ \bibinfo {pages} {5669} (\bibinfo {year}
  {2005})%
  \bibAnnoteFile{NoStop}{degregorio2005}%
\bibitem{adam1965}%
  \BibitemOpen
  \bibfield{author}{%
  \bibinfo {author} {\bibfnamefont{G.}~\bibnamefont{Adam}}\ and\ \bibinfo
  {author} {\bibfnamefont{J.~H.}\ \bibnamefont{Gibbs}},\ }%
  \bibfield{journal}{%
  \bibinfo {journal} {J. Chem. Phys.}\ }%
  \textbf{\bibinfo {volume} {43}},\ \bibinfo {pages} {139} (\bibinfo {year}
  {1965})%
  \bibAnnoteFile{NoStop}{adam1965}%
\bibitem{sastry1998}%
  \BibitemOpen
  \bibfield{author}{%
  \bibinfo {author} {\bibfnamefont{S.}~\bibnamefont{Sastry}}, \bibinfo {author}
  {\bibfnamefont{P.~G.}\ \bibnamefont{Debenedetti}},\ and\ \bibinfo {author}
  {\bibfnamefont{F.~H.}\ \bibnamefont{Stillinger}},\ }%
  \bibfield{journal}{%
  \bibinfo {journal} {Nature}\ }%
  \textbf{\bibinfo {volume} {393}},\ \bibinfo {pages} {554} (\bibinfo {year}
  {1998})%
  \bibAnnoteFile{NoStop}{sastry1998}%
\bibitem{moreno2006}%
  \BibitemOpen
  \bibfield{author}{%
  \bibinfo {author} {\bibfnamefont{A.~J.}\ \bibnamefont{Moreno}}, \bibinfo
  {author} {\bibfnamefont{I.}~\bibnamefont{Saika-Voivod}}, \bibinfo {author}
  {\bibfnamefont{E.}~\bibnamefont{Zaccarelli}}, \bibinfo {author}
  {\bibfnamefont{E.}~\bibnamefont{{La Nave}}}, \bibinfo {author}
  {\bibfnamefont{S.~V.}\ \bibnamefont{Buldyrev}}, \bibinfo {author}
  {\bibfnamefont{P.}~\bibnamefont{Tartaglia}},\ and\ \bibinfo {author}
  {\bibfnamefont{F.}~\bibnamefont{Sciortino}},\ }%
  \bibfield{journal}{%
  \bibinfo {journal} {J. Chem. Phys.}\ }%
  \textbf{\bibinfo {volume} {124}},\ \bibinfo {pages} {204509} (\bibinfo {year}
  {2006})%
  \bibAnnoteFile{NoStop}{moreno2006}%
\bibitem{hoover1968}%
  \BibitemOpen
  \bibfield{author}{%
  \bibinfo {author} {\bibfnamefont{W.~G.}\ \bibnamefont{Hoover}}\ and\ \bibinfo
  {author} {\bibfnamefont{F.~H.}\ \bibnamefont{Ree}},\ }%
  \bibfield{journal}{%
  \bibinfo {journal} {J. Chem. Phys.}\ }%
  \textbf{\bibinfo {volume} {49}},\ \bibinfo {pages} {3609} (\bibinfo {year}
  {1968})%
  \bibAnnoteFile{NoStop}{hoover1968}%
\bibitem{rintoul1996b}%
  \BibitemOpen
  \bibfield{author}{%
  \bibinfo {author} {\bibfnamefont{M.~D.}\ \bibnamefont{Rintoul}}\ and\
  \bibinfo {author} {\bibfnamefont{S.}~\bibnamefont{Torquato}},\ }%
  \bibfield{journal}{%
  \bibinfo {journal} {J. Chem. Phys.}\ }%
  \textbf{\bibinfo {volume} {105}},\ \bibinfo {pages} {9258} (\bibinfo {year}
  {1996})%
  \bibAnnoteFile{NoStop}{rintoul1996b}%
\bibitem{ciamarra2003b}%
  \BibitemOpen
  \bibfield{author}{%
  \bibinfo {author} {\bibfnamefont{M.}~\bibnamefont{{Pica Ciamarra}}}, \bibinfo
  {author} {\bibfnamefont{M.}~\bibnamefont{Tarzia}}, \bibinfo {author}
  {\bibfnamefont{A.}~\bibnamefont{de~Candia}},\ and\ \bibinfo {author}
  {\bibfnamefont{A.}~\bibnamefont{Coniglio}},\ }%
  \bibfield{journal}{%
  \bibinfo {journal} {Phys. Rev. E}\ }%
  \textbf{\bibinfo {volume} {68}},\ \bibinfo {pages} {066111} (\bibinfo {year}
  {2003})%
  \bibAnnoteFile{NoStop}{ciamarra2003b}%
\bibitem{cellai2004}%
  \BibitemOpen
  \bibfield{author}{%
  \bibinfo {author} {\bibfnamefont{D.}~\bibnamefont{Cellai}}, \bibinfo {author}
  {\bibfnamefont{H.}~\bibnamefont{Cuevas}}, \bibinfo {author}
  {\bibfnamefont{A.}~\bibnamefont{Lawlor}}, \bibinfo {author}
  {\bibfnamefont{G.~D.}\ \bibnamefont{McCullagh}},\ and\ \bibinfo {author}
  {\bibfnamefont{K.~A.}\ \bibnamefont{Dawson}},\ }%
  \bibfield{journal}{%
  \bibinfo {journal} {Phys. Rev. E}\ }%
  \textbf{\bibinfo {volume} {70}},\ \bibinfo {pages} {022401} (\bibinfo {year}
  {2004})%
  \bibAnnoteFile{NoStop}{cellai2004}%
\bibitem{marinari2006}%
  \BibitemOpen
  \bibfield{author}{%
  \bibinfo {author} {\bibfnamefont{E.}~\bibnamefont{Marinari}}\ and\ \bibinfo
  {author} {\bibfnamefont{V.}~\bibnamefont{{Van Kerrebroeck}}},\ }%
  \bibfield{journal}{%
  \bibinfo {journal} {Europhys. Lett.}\ }%
  \textbf{\bibinfo {volume} {73}},\ \bibinfo {pages} {383} (\bibinfo {year}
  {2006})%
  \bibAnnoteFile{NoStop}{marinari2006}%
\bibitem{witman2006}%
  \BibitemOpen
  \bibfield{author}{%
  \bibinfo {author} {\bibfnamefont{J.~E.}\ \bibnamefont{Witman}}\ and\ \bibinfo
  {author} {\bibfnamefont{Z.}~\bibnamefont{Wang}},\ }%
  \bibfield{journal}{%
  \bibinfo {journal} {J. Phys. Chem. B}\ }%
  \textbf{\bibinfo {volume} {110}},\ \bibinfo {pages} {6312} (\bibinfo {year}
  {2006})%
  \bibAnnoteFile{NoStop}{witman2006}%
\bibitem{krzakala2008}%
  \BibitemOpen
  \bibfield{author}{%
  \bibinfo {author} {\bibfnamefont{F.}~\bibnamefont{Krzakala}}, \bibinfo
  {author} {\bibfnamefont{M.}~\bibnamefont{Tarzia}},\ and\ \bibinfo {author}
  {\bibfnamefont{L.}~\bibnamefont{Zdeborov\'a}},\ }%
  \bibfield{journal}{%
  \bibinfo {journal} {Physical Review Letters}\ }%
  \textbf{\bibinfo {volume} {101}} (\bibinfo {year} {2008}),\ \doi{\bibinfo
  {doi} {10.1103/PhysRevLett.101.165702}},\
  \url{http://dx.doi.org/10.1103/PhysRevLett.101.165702}%
  \bibAnnoteFile{NoStop}{krzakala2008}%
\bibitem{weeks2000}%
  \BibitemOpen
  \bibfield{author}{%
  \bibinfo {author} {\bibfnamefont{E.~R.}\ \bibnamefont{Weeks}}, \bibinfo
  {author} {\bibfnamefont{J.~C.}\ \bibnamefont{Crocker}}, \bibinfo {author}
  {\bibfnamefont{A.~C.}\ \bibnamefont{Levitt}}, \bibinfo {author}
  {\bibfnamefont{A.}~\bibnamefont{Schofield}},\ and\ \bibinfo {author}
  {\bibfnamefont{D.~A.}\ \bibnamefont{Weitz}},\ }%
  \bibfield{journal}{%
  \bibinfo {journal} {Science}\ }%
  \textbf{\bibinfo {volume} {287}},\ \bibinfo {pages} {627} (\bibinfo {year}
  {2000})%
  \bibAnnoteFile{NoStop}{weeks2000}%
\bibitem{auer2003}%
  \BibitemOpen
  \bibfield{author}{%
  \bibinfo {author} {\bibfnamefont{S.}~\bibnamefont{Auer}}\ and\ \bibinfo
  {author} {\bibfnamefont{D.}~\bibnamefont{Frenkel}},\ }%
  \bibfield{journal}{%
  \bibinfo {journal} {J. Chem. Phys.}\ }%
  \textbf{\bibinfo {volume} {120}},\ \bibinfo {pages} {3015} (\bibinfo {year}
  {2003})%
  \bibAnnoteFile{NoStop}{auer2003}%
\bibitem{pan2005}%
  \BibitemOpen
  \bibfield{author}{%
  \bibinfo {author} {\bibfnamefont{A.~C.}\ \bibnamefont{Pan}}, \bibinfo
  {author} {\bibfnamefont{J.~P.}\ \bibnamefont{Garrahan}},\ and\ \bibinfo
  {author} {\bibfnamefont{D.}~\bibnamefont{Chandler}},\ }%
  \bibfield{journal}{%
  \bibinfo {journal} {Phys. Rev. E}\ }%
  \textbf{\bibinfo {volume} {72}},\ \bibinfo {pages} {041106} (\bibinfo {year}
  {2005})%
  \bibAnnoteFile{NoStop}{pan2005}%
\bibitem{darst2010}%
  \BibitemOpen
  \bibfield{author}{%
  \bibinfo {author} {\bibfnamefont{R.~K.}\ \bibnamefont{Darst}}, \bibinfo
  {author} {\bibfnamefont{D.~R.}\ \bibnamefont{Reichman}},\ and\ \bibinfo
  {author} {\bibfnamefont{G.}~\bibnamefont{Biroli}},\ }%
  \bibfield{journal}{%
  \Doi{10.1063/1.3298877}{\bibinfo {journal} {The Journal of Chemical
  Physics}}\ }%
  \textbf{\bibinfo {volume} {132}},\ \bibinfo {pages} {044510+} (\bibinfo
  {month} {January}\ \bibinfo {year} {2010}),\
  \Eprint{http://arxiv.org/abs/0911.0479}{arXiv:0911.0479},\
  \url{http://dx.doi.org/10.1063/1.3298877}%
  \bibAnnoteFile{NoStop}{darst2010}%
\bibitem{pusey1986}%
  \BibitemOpen
  \bibfield{author}{%
  \bibinfo {author} {\bibfnamefont{P.~N.}\ \bibnamefont{Pusey}}\ and\ \bibinfo
  {author} {\bibfnamefont{W.}~\bibnamefont{van Megen}},\ }%
  \bibfield{journal}{%
  \bibinfo {journal} {Nature}\ }%
  \textbf{\bibinfo {volume} {320}},\ \bibinfo {pages} {340} (\bibinfo {year}
  {1986})%
  \bibAnnoteFile{NoStop}{pusey1986}%
\bibitem{cellai2011}%
  \BibitemOpen
  \bibfield{author}{%
  \bibinfo {author} {\bibfnamefont{D.}~\bibnamefont{Cellai}}, \bibinfo {author}
  {\bibfnamefont{A.~Z.}\ \bibnamefont{Fima}}, \bibinfo {author}
  {\bibfnamefont{A.}~\bibnamefont{Lawlor}},\ and\ \bibinfo {author}
  {\bibfnamefont{K.~A.}\ \bibnamefont{Dawson}}}%
   (\bibinfo {month} {Feb.}\ \bibinfo {year} {2011}),\
  \Eprint{http://arxiv.org/abs/1102.4968}{arXiv:1102.4968},\
  \url{http://arxiv.org/abs/1102.4968}%
  \bibAnnoteFile{NoStop}{cellai2011}%
\bibitem{frenkel2002book}%
  \BibitemOpen
  \bibfield{author}{%
  \bibinfo {author} {\bibfnamefont{D.}~\bibnamefont{Frenkel}}\ and\ \bibinfo
  {author} {\bibfnamefont{B.}~\bibnamefont{Smit}},\ }%
  \emph{\bibinfo {title} {Understanding Molecular Simulation}}\ (\bibinfo
  {publisher} {Academic Press},\ \bibinfo {address} {San Diego, California,
  USA},\ \bibinfo {year} {2002})%
  \bibAnnoteFile{NoStop}{frenkel2002book}%
\bibitem{binder1997}%
  \BibitemOpen
  \bibfield{author}{%
  \bibinfo {author} {\bibfnamefont{K.}~\bibnamefont{Binder}},\ }%
  \bibfield{journal}{%
  \bibinfo {journal} {Rep. Prog. Phys.}\ }%
  \textbf{\bibinfo {volume} {60}},\ \bibinfo {pages} {487} (\bibinfo {year}
  {1997})%
  \bibAnnoteFile{NoStop}{binder1997}%
\bibitem{mccullagh2005}%
  \BibitemOpen
  \bibfield{author}{%
  \bibinfo {author} {\bibfnamefont{G.~D.}\ \bibnamefont{McCullagh}}, \bibinfo
  {author} {\bibfnamefont{D.}~\bibnamefont{Cellai}}, \bibinfo {author}
  {\bibfnamefont{A.}~\bibnamefont{Lawlor}},\ and\ \bibinfo {author}
  {\bibfnamefont{K.~A.}\ \bibnamefont{Dawson}},\ }%
  \bibfield{journal}{%
  \bibinfo {journal} {Phys Rev E}\ }%
  \textbf{\bibinfo {volume} {71}},\ \bibinfo {pages} {030102(R)} (\bibinfo
  {year} {2005})%
  \bibAnnoteFile{NoStop}{mccullagh2005}%
\bibitem{debenedetti1996}%
  \BibitemOpen
  \bibfield{author}{%
  \bibinfo {author} {\bibfnamefont{P.~G.}\ \bibnamefont{Debenedetti}},\ }%
  \emph{\bibinfo {title} {Metastable Liquids}}\ (\bibinfo {publisher}
  {Princeton University Press},\ \bibinfo {address} {Princeton, NJ, USA},\
  \bibinfo {year} {1996})%
  \bibAnnoteFile{NoStop}{debenedetti1996}%
\bibitem{bassler1987}%
  \BibitemOpen
  \bibfield{author}{%
  \bibinfo {author} {\bibfnamefont{H.}~\bibnamefont{B\"assler}},\ }%
  \bibfield{journal}{%
  \Doi{10.1103/PhysRevLett.58.767}{\bibinfo {journal} {Phys. Rev. Lett.}}\ }%
  \textbf{\bibinfo {volume} {58}},\ \bibinfo {pages} {767} (\bibinfo {month}
  {Feb}\ \bibinfo {year} {1987})%
  \bibAnnoteFile{NoStop}{bassler1987}%
\bibitem{kegel2000}%
  \BibitemOpen
  \bibfield{author}{%
  \bibinfo {author} {\bibfnamefont{W.~K.}\ \bibnamefont{Kegel}}\ and\ \bibinfo
  {author} {\bibfnamefont{A.}~\bibnamefont{van Blaaderen}},\ }%
  \bibfield{journal}{%
  \bibinfo {journal} {Science}\ }%
  \textbf{\bibinfo {volume} {287}},\ \bibinfo {pages} {290} (\bibinfo {year}
  {2000})%
  \bibAnnoteFile{NoStop}{kegel2000}%
\bibitem{toninelli2005}%
  \BibitemOpen
  \bibfield{author}{%
  \bibinfo {author} {\bibfnamefont{C.}~\bibnamefont{Toninelli}}, \bibinfo
  {author} {\bibfnamefont{G.}~\bibnamefont{Biroli}},\ and\ \bibinfo {author}
  {\bibfnamefont{D.~S.}\ \bibnamefont{Fisher}},\ }%
  \bibfield{journal}{%
  \Doi{10.1007/s10955-005-5250-z}{\bibinfo {journal} {Journal of Statistical
  Physics}}\ }%
  \textbf{\bibinfo {volume} {120}},\ \bibinfo {pages} {167} (\bibinfo {month}
  {July}\ \bibinfo {year} {2005}),\ ISSN \bibinfo {issn} {0022-4715},\
  \url{http://dx.doi.org/10.1007/s10955-005-5250-z}%
  \bibAnnoteFile{NoStop}{toninelli2005}%
\bibitem{glotzer2000b}%
  \BibitemOpen
  \bibfield{author}{%
  \bibinfo {author} {\bibfnamefont{S.~C.}\ \bibnamefont{Glotzer}}, \bibinfo
  {author} {\bibfnamefont{N.}~\bibnamefont{Jan}},\ and\ \bibinfo {author}
  {\bibfnamefont{P.~H.}\ \bibnamefont{Poole}},\ }%
  \bibfield{journal}{%
  \Doi{10.1088/0953-8984/12/29/337}{\bibinfo {journal} {Journal of Physics:
  Condensed Matter}}\ }%
  \textbf{\bibinfo {volume} {12}},\ \bibinfo {pages} {6675+} (\bibinfo {month}
  {July}\ \bibinfo {year} {2000}),\ ISSN \bibinfo {issn} {0953-8984},\
  \url{http://dx.doi.org/10.1088/0953-8984/12/29/337}%
  \bibAnnoteFile{NoStop}{glotzer2000b}%
\bibitem{simeonova2004}%
  \BibitemOpen
  \bibfield{author}{%
  \bibinfo {author} {\bibfnamefont{N.~B.}\ \bibnamefont{Simeonova}}\ and\
  \bibinfo {author} {\bibfnamefont{W.~K.}\ \bibnamefont{Kegel}},\ }%
  \bibfield{journal}{%
  \bibinfo {journal} {Phys. Rev. Lett.}\ }%
  \textbf{\bibinfo {volume} {93}},\ \bibinfo {pages} {035701} (\bibinfo {year}
  {2004})%
  \bibAnnoteFile{NoStop}{simeonova2004}%
\bibitem{yu2009}%
  \BibitemOpen
  \bibfield{author}{%
  \bibinfo {author} {\bibfnamefont{Y.}~\bibnamefont{Yu}}, \bibinfo {author}
  {\bibfnamefont{S.~M.}\ \bibnamefont{Anthony}}, \bibinfo {author}
  {\bibfnamefont{S.~C.}\ \bibnamefont{Bae}}, \bibinfo {author}
  {\bibfnamefont{E.}~\bibnamefont{Luijten}},\ and\ \bibinfo {author}
  {\bibfnamefont{S.}~\bibnamefont{Granick}},\ }%
  \bibfield{journal}{%
  \Doi{10.1080/15421400903048024}{\bibinfo {journal} {Molecular Crystals and
  Liquid Crystals}}\ }%
  \textbf{\bibinfo {volume} {507}},\ \bibinfo {pages} {18} (\bibinfo {year}
  {2009}),\ \url{http://dx.doi.org/10.1080/15421400903048024}%
  \bibAnnoteFile{NoStop}{yu2009}%
\bibitem{terraneo2009}%
  \BibitemOpen
  \bibfield{author}{%
  \bibinfo {author} {\bibfnamefont{F.~M.}\ \bibnamefont{Terraneo}}, \bibinfo
  {author} {\bibfnamefont{P.}~\bibnamefont{Pinto}},\ and\ \bibinfo {author}
  {\bibfnamefont{K.~A.}\ \bibnamefont{Dawson}},\ }%
  \bibfield{journal}{%
  \Doi{10.1209/0295-5075/85/26002}{\bibinfo {journal} {EPL (Europhysics
  Letters)}}\ }%
  \textbf{\bibinfo {volume} {85}},\ \bibinfo {pages} {26002+} (\bibinfo {year}
  {2009}),\ \url{http://dx.doi.org/10.1209/0295-5075/85/26002}%
  \bibAnnoteFile{NoStop}{terraneo2009}%
\bibitem{weigt2003}%
  \BibitemOpen
  \bibfield{author}{%
  \bibinfo {author} {\bibfnamefont{M.}~\bibnamefont{Weigt}}\ and\ \bibinfo
  {author} {\bibfnamefont{A.~K.}\ \bibnamefont{Hartmann}},\ }%
  \bibfield{journal}{%
  \bibinfo {journal} {Europhys. Lett.}\ }%
  \textbf{\bibinfo {volume} {64}},\ \bibinfo {pages} {533} (\bibinfo {year}
  {2003})%
  \bibAnnoteFile{NoStop}{weigt2003}%
\end{thebibliography}

%

\end{document}